\documentclass[twocolumn,superscriptaddress,pr]{revtex4}
\usepackage{verbatim}
\usepackage{amsmath,amssymb}
\usepackage{graphicx}
\usepackage{color}
\usepackage[colorlinks,bookmarks=false,citecolor=blue,
linkcolor=red,urlcolor=blue]{hyperref}
\usepackage{times}

\begin{document}

\author{Vincenzo Alba}
\affiliation{Department of Physics and Arnold Sommerfeld
Center for Theoretical Physics, Ludwig-Maximilians-Universit\"at
M\"unchen, D-80333 M\"unchen, Germany}

\author{Fabian Heidrich-Meisner}
\affiliation{Department of Physics and Arnold Sommerfeld
Center for Theoretical Physics, Ludwig-Maximilians-Universit\"at
M\"unchen, D-80333 M\"unchen, Germany}

\date{\today}

\title{Entanglement spreading after a geometric quench in quantum spin chains}

\begin{abstract} 
We investigate the entanglement spreading in the anisotropic spin-$1/2$ Heisenberg ($XXZ$) 
chain after a {\it geometric quench}. This corresponds to a sudden change of the 
geometry of the chain or, in the equivalent language of interacting fermions confined in a 
box trap, to a sudden increase of the trap size. 
The entanglement dynamics after the quench is associated with the ballistic propagation of a 
magnetization wavefront. At the free fermion point ($XX$ chain), the von Neumann entropy $S_A$ 
exhibits several intriguing dynamical regimes. Specifically, at short times a logarithmic 
increase is observed, similar to local quenches. This is accurately described by an analytic 
formula that we derive from heuristic arguments. At intermediate times partial revivals of the 
short-time dynamics are superposed with a power-law increase $S_A\sim t^{\alpha}$, with $\alpha<1$. 
Finally, at very long times a steady state develops with constant entanglement entropy, 
apart from oscillations. As expected, 
since the model is integrable, we find that the steady state is  {\it non} thermal, although it 
exhibits {\it extensive} entanglement entropy. We also investigate the entanglement dynamics 
after the quench from a finite to the infinite chain (sudden expansion). While at long times the 
entanglement vanishes, we demonstrate that its relaxation dynamics exhibits a number of  
scaling properties. Finally, we discuss the short-time entanglement dynamics in the $XXZ$ chain in the 
gapless phase. The same formula that describes the time dependence for the  $XX$ chain 
remains valid in the whole gapless phase. 
\end{abstract}


\maketitle

\section{Introduction}

The recent extraordinary progress achieved with trapped cold-atomic gases 
experiments has boosted a renewed theoretical interest in the 
dynamics of isolated quantum many-body systems out-of-equilibrium~\cite{polkovnikov-2011,bloch-2008,rigol-2008}. 
Highly-investigated topics include the relaxation dynamics~\cite{greiner-2002,hofferberth-2007,trotzky-2012,
cheneau-2012,gring-2012,meinert-2013}, thermalization~\cite{kinoshita-2006,hofferberth-2007} in 
out-of-equilibrium steady states, and transport properties~\cite{schneider-2012,ronzheimer-2013,reinhard-2013}. 
A popular out-of-equilibrium experiment is the so-called {\it quantum quench}, 
in which a system is initially prepared in the ground state of a many-body quantum 
Hamiltonian, and  a non-trivial unitary dynamics is then induced by changing 
instantaneously (i.e., ``quenching'') one (or many) control parameters. Depending 
on whether this change happens locally or in the whole system, the quench falls 
into the class of local or global quenches, respectively. 

Entanglement measures are nowadays accepted as useful tools to extract universal 
properties of quantum many-body systems, both in and out-of-equilibrium~\cite{holzhey-1994,
vidal-2003,latorre-2004,calabrese-2004,amico-2008,calabrese-2009-rev,eisert-2010}. 
Considering a bipartition of a system that is in a pure state $|\psi\rangle$ into 
parts $A$ and $B$, a standard measure of their mutual entanglement is the von Neumann 
entropy $S_A$ 
\begin{equation}  
\label{entropy}
S_A=-\textrm{Tr}\rho_A\log\rho_A. 
\end{equation}
Here $\rho_A$ is the reduced density matrix for $A$, obtained after tracing 
part $B$ from the full density matrix $\rho\equiv |\psi\rangle\langle\psi|$.

The real-time entanglement dynamics (and that of related quantities) after a quantum quench has been intensively  
investigated in recent years, both analytically (using conformal field 
theory~\cite{calabrese-2007,calabrese-2007a,stefan-2011} and for exactly 
solvable models~\cite{eisler-2007,eisler-2008,fagotti-2008,igloi-2009,hsu-2009,divakaran-2011,igloi-2012,schoenhammer,levine-2012,sabetta-2013,eisler-2013,collura-2014,bucciantini-2014}) 
and numerically~\cite{de-chiara-2006,lauchli-2008,eisler-2012,kim-2013,collura-2013,tagliacozzo-2013,schaenmayer-2013,kessler-2013,collura-2014a,zamora-2014}. 
The nature of the quench is strikingly reflected in the time dependence of 
the von Neumann entropy: while local quenches are associated with a logarithmic growth~\cite{eisler-2007,calabrese-2007,
stefan-2011}, 
a more dramatic (linear) behavior is observed in global quenches~\cite{calabrese-2005,de-chiara-2006,calabrese-2007a,lauchli-2008}. 
This is related to the different excess energy density, measured with respect to the ground-state energy of the 
post-quench Hamiltonian. This excess energy remains finite in the global quench protocol, whereas 
it vanishes in the local one, upon increasing the system size. 
We mention that a scheme  for measuring entanglement dynamics in cold-atomic gases experiments has recently 
been proposed in Ref.~\onlinecite{daley-2012,abanin-2012}. 

\begin{figure}
\includegraphics[width=.95\linewidth]{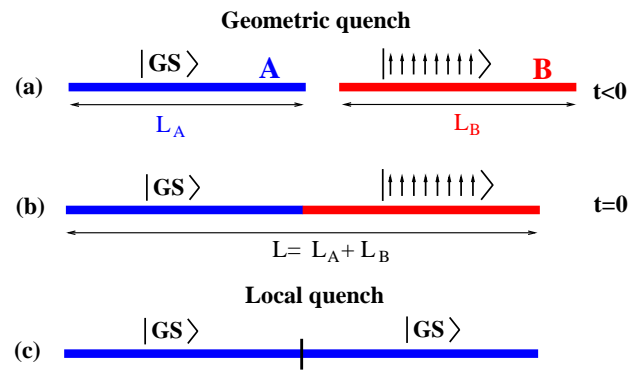}
\caption{({a}), ({b}) Geometric quench in $XXZ$ spin chain: quench setup 
 for open boundary conditions. ({a}) $t<0$: Two independent chains ($A$ 
 and $B$) of length $L_A$ and $L_B\equiv L-L_A$ are prepared 
 in the ground state of the $XXZ$ chain in the sectors with zero and maximal 
 magnetization, respectively. ({b}) $t=0$: $A$ and $B$ are glued 
 together. Here we consider geometric quenches with several aspect 
 ratios $\omega\equiv L_A/L$, i.e., $0\le\omega\le 1$, focusing 
 on the entanglement between $A$ and $B$. ({c}) Example of {\it typical}  
 {\it local} quench protocol: the initial state at $t=0$ is obtained connecting 
 two {\it identical} chains prepared in the ground state at zero magnetization. 
}
\label{fig00:quench_prot}
\end{figure}

In our work we focus on a situation that is intermediate between a local and a global 
quench, considering the real-time entanglement dynamics following an instantaneous 
change of the geometry or the size of the system, the so-called {\it geometric quench}, as discussed 
in Ref.~\onlinecite{mossel-2010}. To be specific, we study the spin-$1/2$ $XXZ$ chain in the gapless 
phase. The quench protocol is described as follows (cf.~Fig.~\ref{fig00:quench_prot}): 
initially two chains $A$ and $B$ are prepared in the ground state of the $XXZ$ model in 
the sector with zero and maximum magnetization (i.e., fully polarized), respectively. 
The unitary dynamics under the $XXZ$ Hamiltonian is then induced by connecting the two chains. 
Alternatively, after mapping the $XXZ$ chain onto a system of interacting fermions confined in 
a box trap (i.e., chain $A$), the geometric quench is equivalent to suddenly increasing the 
trap size. Notice that this is similar to the so-called sudden expansion protocol  
used in cold-atomic gases experiments~\cite{kinoshita-2004,schneider-2012,ronzheimer-2013,reinhard-2013}, 
in which particles are released from the trap and expand in an empty  lattice. This sudden 
expansion has been studied theoretically in, e.g., Refs.~\onlinecite{rigol-2004,rigol-2005,rigol-2005a,minguzzi-2005,gangardt-2008,fabian-2008,fabian-2009,langer-2012,vidmar-2013,
kajala-2013,kessler-2013}.

Clearly, as the two chains are prepared in their respective ground states, the post-quench 
dynamics is induced by a ``defect'' at the interface between $A$ and $B$, which is a distinctive 
feature of local quenches. On the other hand, the excess energy density is {\it finite}, as 
in global quenches. Notice that the initial state after the quench is of the ``domain wall'' 
type (i.e., spatially inhomogeneous), and the ensuing out-of-equilibrium dynamics 
has been at the focus of many recent theoretical studies~\cite{antal-1999,karevski-2002,gobert-2005,steiniwegeg-2006,santos-2008,haque-2010,lancaster-2010,lancaster-2010a,mossel-2010a,jesenko-2011,santos-2011,eisler-2013,halimeh-2013,sabetta-2013}.
For instance, the state $|m_A\rangle\otimes |m_B\rangle$, with $m_A$ and $m_B$ 
being the total magnetization in chain $A$ and $B$, respectively, provides a basic setup for 
studying transport-related questions and non-equilibrium steady-state
properties such as the conditions for ballistic or diffusive dynamics in integrable many-body systems in one 
dimension~\cite{zotos-1997,heidrich-meisner-2007}). 
In particular, the sub-class of initial states with $m_A=-m_B\equiv m$ has been extensively  
studied~\cite{antal-1999,gobert-2005,steiniwegeg-2006,santos-2008,santos-2011,lancaster-2010,lancaster-2010a,eisler-2013,halimeh-2013,sabetta-2013}. 
The initial state in our work corresponds to choosing $m_A=0$ and $m_B=L_B/2$. Interestingly, in the 
situation with $m_A=-m_B$ it has been found that the magnetization dynamics during the domain wall melting 
is ballistic close to the free fermion point, super-diffusive at the isotropic point, and diffusive in the 
gapped phases \cite{gobert-2005}. While entanglement 
dynamics from domain wall initial states is interesting as such, it is also important for the simulability of 
quench dynamics using matrix-product states based methods, such as DMRG (Density Matrix 
Renormalization Group)~\cite{white-1992,schollwoeck-2005,schollwoeck-2011}. For a discussion of transport and local 
quenches in spin chains in non-equilibrium for other initial conditions, see Ref.~\cite{langer-2009,langer-2011,steiniwegeg-2010,ganahl-2012,fukuhara-2013,fukuhara-2013a,zauner-2012,karrasch-2014}. We should also mention that transport and entanglement properties have also been studied in the dynamics induced by local 
impurities~\cite{igloi-2009,igloi-2012,eisler-2012,collura-2013,igloi-2014}.

\paragraph*{Summary of the results.---}
In this work we fully characterize the entanglement spreading after a generic 
geometric quench, focusing on the entanglement entropy between $A$ and $B$. 
The spreading of information (and the related entanglement increase) is 
associated with the propagation of an extended magnetization wavefront. The two 
edges of the front expand {\it ballistically} 
in the $A$ and $B$ parts of the chain, with two different velocities. These coincide at 
the free-fermion point ($XX$ chain, i.e., vanishing anisotropy), where the wavefront  
propagates symmetrically. For the $XX$ chain the full magnetization profile, at any time 
after the quench, is obtained analytically, using a semiclassical reasoning and 
free-fermionic techniques (as in~\cite{antal-1997}). For the $XXZ$ chain, although we do 
not derive analytically the full magnetization profile, we provide an approximate expression 
describing the central region of the wavefront. 

The entanglement evolution exhibits several dynamical behaviors at different 
time scales. For the $XX$ chain all these dynamical regimes are thoroughly investigated,  
exploiting the mapping to free fermions. At {\it short} times,  the von Neumann 
entropy increases logarithmically, as in a local quench. Although the well-known 
conformal field theory (CFT) result for the local quench~\cite{calabrese-2007,stefan-2011} 
does not apply, we provide a heuristic extension of this result to our case, which accurately 
reproduces the entanglement dynamics. One remarkable consequence is that the entanglement 
dynamics, apart from a size-dependent shift, is described by a scaling function $f_s(y)$,  
with $y\equiv t/L_A$ ($t$ is the time after the quench and $L_A$ the size of part $A$). 
We numerically demonstrate that the same scaling holds true in the 
interacting case.

At {\it intermediate} times the entanglement entropy exhibits revivals of 
the short-time dynamics, superposed with a power-law increase as $S_A\sim t^\alpha$ 
(apart from possible multiplicative logarithmic corrections). We numerically extract 
the exponent $\alpha$, finding $\alpha<1$. This suggests that the geometric quench 
cannot be thought of as a simple superposition of a logarithmic (i.e., local-quench 
like) and a linear (as in global quenches) behaviors. 

At {\it long} times the system reaches an out-of-equilibrium steady state, and 
the entropy oscillates around a stationary value. The steady state is {\it non} thermal 
and shows features of the initial Fermi surface in part $A$ of the chain. Finally, in 
spite of the non-thermal nature of the steady state, we demonstrate that the entanglement 
entropy is {\it extensive}, and its stationary value can be determined analytically. Similar 
results have been found in Ref.~\onlinecite{sabetta-2013} for the quench from the state $|-m\rangle\otimes|
m\rangle$, while the constraints put on steady states due to integrability for a similar 
set-up and hard-core bosons, which map on the $XX$ model, were discussed 
in a seminal paper by Rigol et al.~\cite{rigol-2007}.

We also discuss the information spreading after the quench from a finite to the infinite chain. 
In the framework of trapped interacting fermions, this corresponds to removing the trap 
completely, i.e., to the so-called sudden expansion. While the entanglement entropy vanishes 
asymptotically (i.e., at large times), its relaxation dynamics shows unexpected scaling behaviors. 
In particular, the entanglement dynamics is described by a scaling function $f_\ell(z)$ with 
$z\equiv t/L_A^2$. Furthermore, $f_\ell(z)$ exhibits an intriguing structure: while at $z\ll 1$ 
one has $f_\ell(z)\sim -\log(1/z)$, a crossover to the behavior $1/z(1-\log(1/z))$ occurs at 
$z\sim 1$. Similar scaling behaviors have been observed in the entanglement dynamics of 
non-interacting fermions in continuous space released from a trap~\cite{calabrese-2011,calabrese-2011a,vicari-2012}.

Finally, by means of tDMRG~\cite{daley-2004,feiguin-2004} (time-dependent 
Density Matrix Renormalization Group) simulations, we investigate the role of interactions on the 
short-time entanglement dynamics, focusing on the $XXZ$ spin chain. Our main result is that the 
same formula used for the $XX$ chain remains valid. Interestingly, since the excitations 
forming the wavefront propagate in the two parts of the chain with different velocities, we find 
that the entanglement spreading rate is not a trivial function of the spinon velocity. 

\paragraph*{Outline.---} 
This paper is organized as follows. In Sec.~\ref{model_protocol} 
we introduce the $XXZ$ spin chain and the geometric quench protocol. Sections~\ref{wave},~\ref{overview},~\ref{short-time},~\ref{steady-state}, and~\ref{inf_size_quench} are devoted to the $XX$ chain. 
In Sec.~\ref{wave} we investigate the magnetization wavefront expansion. An overview of the entanglement 
dynamics after the quench is given in Sec.~\ref{overview}, while the short-time behavior is discussed 
in detail in Sec.~\ref{short-time}. In Sec~\ref{steady-state} we characterize the entanglement 
properties in the steady state. The infinite-chain quench is then discussed in Sec.~\ref{inf_size_quench}. 
Finally, in Sec.~\ref{gq_xxz} we investigate the short-time entanglement dynamics in the $XXZ$ model, while a summary is 
provided in Sec.~\ref{summary}.

\section{Model \& quench protocol}
\label{model_protocol}

\subsection{The spin-$1/2$ $XXZ$ spin chain}
The open $XXZ$ spin chain of length $L$ is defined by the Hamiltonian
\begin{equation}
{\mathcal H}=\frac{J}{2}\sum\limits_{i=1}^{L-1}(S_i^+S_{i+1}^-+h.c.) 
+J\Delta\sum\limits_{i=1}^{L-1}S_i^zS_{i+1}^z.
\label{ham-xxz}
\end{equation}
Here $S_i^\pm\equiv S_i^x\pm iS_i^y$, $S_i^z\equiv S^z_i$   
are spin-$1/2$ operators acting at site $i$ of the chain, and $\Delta$ the 
so-called anisotropy (we set $J=1$ in Eq.~\eqref{ham-xxz}). For a periodic chain  
an extra term in Eq.~\eqref{ham-xxz} connects sites $1$ and $L$. 
The ground-state phase diagram of the $XXZ$ chain exhibits a gapless spin-liquid 
phase at $-1<\Delta\le 1$, while it is gapped at $|\Delta|>1$. At $\Delta=0$ 
($XX$ chain) the $XXZ$ chain reduces to a free-fermionic model (cf. Appendix~\ref{xx-chain} 
for more details)~\cite{mikeska-2004}.

The low-energy spectrum of Eq.~\eqref{ham-xxz} is linear in the spin liquid phase, 
and it is described (along with other low-energy properties) by a conformal 
field theory (CFT) with central charge $c=1$. At sufficiently large $L$ one 
has~\cite{cardy-1986,affleck-1986}
\begin{equation}
E_\alpha=LE_{\rm bulk}+E_{\rm bound}+\frac{\pi v_s}{L}\left(h_\alpha-
\frac{c}{24}\right)+{\mathcal O}(L^{-2}), 
\label{cft-spectrum}
\end{equation}
with $E_\alpha$ being the energy of a low-lying excitation (labeled by $\alpha\in
{\mathbb N}$) of Eq.~\eqref{ham-xxz}. In Eq.~\eqref{cft-spectrum}, $E_{\rm bulk}$ and 
$E_{\rm bound}$ are the usual bulk (extensive) and a boundary (in presence 
of non-periodic boundary conditions) contributions, $c$ is the central charge (here $c=1$), 
and~$v_s$ the spinon velocity. Finally, $h_\alpha$ are the scaling dimensions of the operators 
(both {\it primary} operators and their {\it descendants}~\cite{yellow-book,ginsparg-cft}) 
appearing in the CFT. In particular, $\alpha=0$ (with $h_\alpha=0$) corresponds to 
the ground-state energy $E_0$. Finite-size deviations from the linear dispersion are 
accounted for by the ${\mathcal O}(1/L^2)$ term. Notice that Eq.~\eqref{cft-spectrum} 
can be thought of as the spectrum of an effective Hamiltonian ${\mathcal H}_{CFT}$. The 
two energy scales set by the terms $\sim 1/L$ and $\sim1/L^2$ in Eq.~\eqref{cft-spectrum} 
imply the existence of two typical time scales $t_s^*\sim L/v_s$ (short times) and 
$t_\ell^*\sim L^2/v_s$ (long times) (here all lengths are measured in units of the 
lattice constant $a=1$). We anticipate that the existence of $t_s^*$ and $t^*_\ell$ 
will be strikingly reflected in the entanglement dynamics after the quench.

\subsection{Quench protocol and observables}

The geometric quench protocol for the $XXZ$ spin chain with open boundary conditions 
(the generalization to periodic boundary conditions is straightforward) is depicted in 
Fig.~\ref{fig00:quench_prot}. At time $t<0$ (Fig.~\ref{fig00:quench_prot} ({a})) two 
disconnected chains $A$ and $B$ (of respective lengths $L_A$ and $L_B$) are prepared in 
the ground state $|GS\rangle$ of Eq.~\eqref{ham-xxz} and in the fully polarized (ferromagnetic) 
state $|F\rangle\equiv|\uparrow\uparrow\cdots\uparrow\rangle$, respectively. The latter 
is an eigenstate of Eq.~\eqref{ham-xxz} at any $\Delta$, with eigenenergy $E\equiv\langle{
\mathcal H}\rangle=(L-1)\Delta/4$. At $\Delta\gg -1$ in the gapless phase, which is the 
region of interest here, $|F\rangle$ is in the high-energy part of the spectrum 
of Eq.~\eqref{ham-xxz}, and at the isotropic point ($\Delta=1$) it is the 
highest-energy eigenstate. 

At $t=0$ the two chains $A$ and $B$ are connected to form a new one of total length 
$L\equiv L_A+L_B$ (cf. Fig.~\ref{fig00:quench_prot} ({b})). The initial quantum 
state $|\Psi_{\rm init}\rangle$ after the quench exhibits a step-like (or ``domain wall'') 
magnetization profile (with $\langle S^z_i\rangle=0$ and $\langle S_i^z\rangle=1/2$ 
for $i\in A$ and $i\in B$ respectively). It is useful to introduce the aspect ratio 
$0\le\omega\le 1$ as 
\begin{equation}
\omega\equiv \frac{L_A}{L}.
\end{equation}
Finally, at $t>0$ the chain evolves unitarily under Eq.~\eqref{ham-xxz} since  
$|\Psi_{\rm init}\rangle$ is not an eigenstate. In this work we focus 
on the real-time dynamics of the von Neumann entropy $S_A$ between $A$ and $B$.

\subsection{Geometric vs local quench}

It is interesting to compare the geometric quench with a {\it local} quench~\cite{calabrese-2007,
eisler-2007,eisler-2007a,peschel-2009,igloi-2009,gobert-2005,divakaran-2011,stefan-2011}. A 
typical local quench is illustrated in Fig.~\ref{fig00:quench_prot} ({c}): the initial state 
$|\Psi_{\rm init}\rangle$ at $t=0$ is now obtained by ``gluing'' together two {\it identical} 
copies of the ground state as $|\Psi_{\rm init}\rangle\equiv|GS\rangle\otimes|GS\rangle$ (which 
implies that $\omega=1/2$).

Clearly, the excess energy density, which is defined as $\delta_e\equiv|\langle\Psi_{\rm init}|
{\mathcal H}|\Psi_{\rm init}\rangle-E_0|/L$, vanishes in the local quench ($\delta_e\sim 
{\mathcal O}(1/L)$, cf. Eq.~\eqref{cft-spectrum}) in the limit $L\to\infty$. Oppositely, in 
the geometric quench, due to the typically large energy of chain $B$, one has $\delta_e\sim{
\mathcal O}(1)$. As a consequence, while only few low-lying excitations (cf. Eq.~\eqref{ham-xxz}) 
play a role in the dynamics after a local quench, this is certainly different in the geometric 
quench. 

In the CFT framework the initial quantum state $|\Psi_{\rm init}\rangle$ can be decomposed 
(in analogy with the standard decomposition in the eigenbasis of Eq.~\eqref{ham-xxz}) as 
\begin{align}
|\Psi_{\rm init}\rangle=\sum\limits_{a}c_a|\phi_a\rangle
\label{cft-dec}
\end{align}
where the sum runs over both primary and descendants fields $\phi_a$ of the CFT. 
In principle Eq.~\eqref{cft-dec} provides all the necessary information about 
the post-quench dynamics, after time evolving each eigenstate $|\phi_a\rangle$ of 
${\mathcal H}_{CFT}$ with $e^{-i{\mathcal H}_{CFT}t}$. 

However, the coefficients $c_a$ in Eq.~\eqref{cft-dec} are not easy to calculate 
for a generic initial state. For the local quench this is possible because 
only one operator (the identity) and its descendants enter in the 
expansion Eq.~\eqref{cft-dec}~\cite{stefan-2011}. This is related to the 
fact that $|\Psi_{\rm init}\rangle$ has substantial overlap only with the ground state 
of Eq.~\eqref{ham-xxz}. A prominent consequence is that the entanglement entropy 
dynamics after a local quench shows perfect revivals (apart from scaling 
corrections) for $t\sim nt_s^*,\,\, n\in{\mathbb N}$, at 
least up to the time $t\sim t_\ell^*$, at which the CFT description is no longer valid, 
cf. Eq.~\eqref{cft-spectrum}~\cite{stefan-2011}. Conversely, this will be strikingly 
different for the geometric quench (cf. Sec.~\ref{overview}).

\subsection{Entanglement dynamics after a local quench}
Here we briefly review the CFT result for the entanglement dynamics after the 
local quench in Fig.~\ref{fig00:quench_prot} ({c})~\cite{calabrese-2007,
stefan-2011}. The real-time dynamics of the von Neumann entropy depends only on 
the central charge $c$, the boundary conditions, and the spinon velocity $v_s$. 
The result reads~\cite{igloi-2009,stefan-2011}
\begin{equation}
\label{loc-cft}
S_A(t)=\nu\frac{c}{3}\log\left|\frac{L_A}{\pi}\sin
\frac{\nu\pi v_st}{2L_A}\right|+k_\nu. 
\end{equation}
Here $\nu=1,2$ are for open (obc) and periodic boundary conditions (pbc), respectively, 
and $k_\nu$ is a non-universal constant. The analog of Eq.~\eqref{loc-cft} for 
aspect ratios $\omega\ne 1/2$ is also known~\cite{stefan-2011}. 
It is useful to rewrite Eq.~\eqref{loc-cft} as 
\begin{equation}
S_A(t)=\nu\frac{c}{3}\log\left|\sin(\pi y)\right|+
\nu\frac{c}{3}\log\left (\frac{L_A}{\pi}\right )+k_\nu. 
\label{loc-cft-1}
\end{equation}
Here $y$ is the rescaled time $y\equiv\nu\pi v_st/(2L_A)$. 
In Eq.~\eqref{loc-cft-1} it is apparent that the entropy dynamics is 
described by a scaling function of $y$, apart from the shift $\nu 
c/3\log(L_A)$. Interestingly, the latter resembles the {\it equilibrium} 
ground-state entropy for a block of size $L_A$ embedded in an infinite 
chain~\cite{holzhey-1994,calabrese-2004,vidal-2003,latorre-2004}. A 
similar scaling will hold for the entanglement dynamics at short times 
after the geometric quench (cf. Sec.~\ref{short-time}).
In the limit $t/L_A\ll 1$ (short times), Eq.~\eqref{loc-cft} reduces 
to~\cite{calabrese-2007}
\begin{equation}
S_A(t)=\nu\frac{c}{3}\log t+k_\nu \,.
\label{loc-asy}
\end{equation}
A similar logarithmic behavior (as in Eq.~\eqref{loc-cft} and Eq.~\eqref{loc-asy}) 
has been observed in the entanglement dynamics induced by local 
impurities or perturbations in spin or particle densities~\cite{collura-2013,igloi-2009,igloi-2012,langer-2011}.

\begin{figure}
\includegraphics[width=1\linewidth]{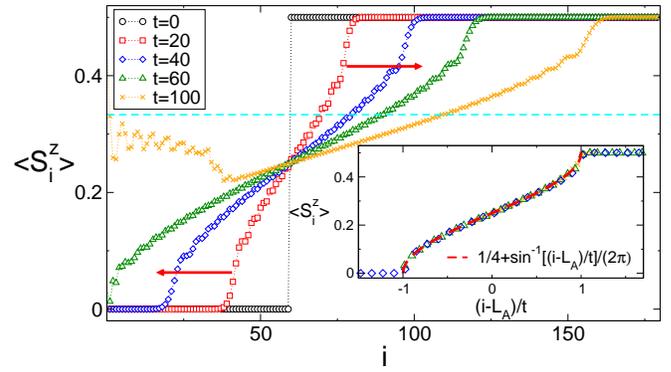}
\caption{ Magnetization wavefront after the geometric 
 quench with aspect ratio $\omega\equiv L_A/L=1/3$ (cf. 
 Fig.~\ref{fig00:quench_prot}) in the open $XX$ chain: 
 local magnetization $\langle S^z_i\rangle$ at 
 site $0\le i<3L_A$ in the chain. Data are exact 
 results for $L_A=60$ and several times. At $t=0$ a 
 step-like profile is present. At $t>0$ a wavefront forms 
 propagating symmetrically (with $v_s=1$) in parts $A$ and $B$ 
 (see horizontal arrows). The dashed line is the  
 flat profile expected on average at $t\to\infty$. At $t=L_A/v_s$ 
 the wavefront is reflected at the (left) boundary of the chain. 
 Inset: rescaled dynamics, $\langle S_i^z\rangle$ versus 
 $(i-L_A)/t$. All data for different times collapse on the 
 same scaling function (dashed line).
 }
\label{fig3:dw_XX}
\end{figure}

\section{Magnetization wavefront after the quench}
\label{wave}

In this section we discuss the real-time dynamics of the magnetization 
profile $\langle S^z_i\rangle$ after a geometric quench. Here $\langle\cdot
\rangle$ denotes the expectation value with respect to the post-quench 
wavefunction. We focus on the open $XX$ chain ($\Delta=0$ in Eq.~\eqref{ham-xxz}). 
At any $i,t$, $\langle S_i^z\rangle$ can be computed analytically 
exploiting the mapping to free fermions (see Appendix~\ref{xx-chain}).

Figure~\ref{fig3:dw_XX} shows $\langle S_i^z\rangle$ versus $1\le i\le L$ at 
several times (denoted by different symbols in the figure) after the geometric quench with 
$\omega\equiv L_A/L=1/3$ and fixed $L_A=60$. At $t=0$, a domain-wall profile 
is present. A magnetization wavefront develops at $t>0$ (as the domain wall 
``melts'') with its left and right edges propagating {\it ballistically} with 
the same velocity $v=v_s=1$, with $v_s$ being the spinon velocity, in part $A$ of the 
chain. At $t=L_A/v_s$, a perfect reflection of the left wavefront edge occurs at the left 
boundary of the chain. Finally, at large times translational invariance is restored, and a 
stationary behavior sets in with uniform magnetization (dashed-horizontal line in the figure).

The ballistic nature of the wavefront dynamics is further supported by the data shown in the inset of 
Fig.~\ref{fig3:dw_XX}, where we plot $\langle S_i^z\rangle$ versus the rescaled variable 
$(i-L_A)/t$. Remarkably, all  data at different times and positions collapse on 
the same scaling curve. This curve can be obtained analytically using a semiclassical 
reasoning that was also applied  in~Refs.~\cite{antal-1997,antal-1999,gobert-2005,antal-2008,eisler-2013,
sabetta-2013} for the quench with initial state $|-m\rangle\otimes|m\rangle$. 
The result reads 
\begin{equation}
\langle S_i^z(t)\rangle=\frac{1}{4}+\\\frac{1}{2\pi}\sin^{-1}
\left[\frac{i-L_A}{v_st}\right], 
\label{XX_wf_theo}
\end{equation}
and is included in Fig.~\ref{fig3:dw_XX} as dashed line. Interestingly, the central 
region of the profile, at $|(i-L_A)/(v_st)|\ll 1$, shows a linear dependence on $(i-L_A)/t$:   
\begin{equation}
\langle S_i^z(t)\rangle\approx \frac{1}{4}+\frac{i-L_A}{v_st}. 
\label{wf_app}
\end{equation}
It is natural to expect that Eq.~\eqref{wf_app} remains valid in the interacting 
case (i.e., nonzero anisotropy), after taking into account the renormalization, due to interactions,
of the the spinon velocity $v_s$ (see Sec.~\ref{wave_int} for a numerical check of Eq.~\eqref{wf_app} 
in the $XXZ$ spin chain).

\begin{figure*}[t]
\includegraphics[width=1\linewidth]{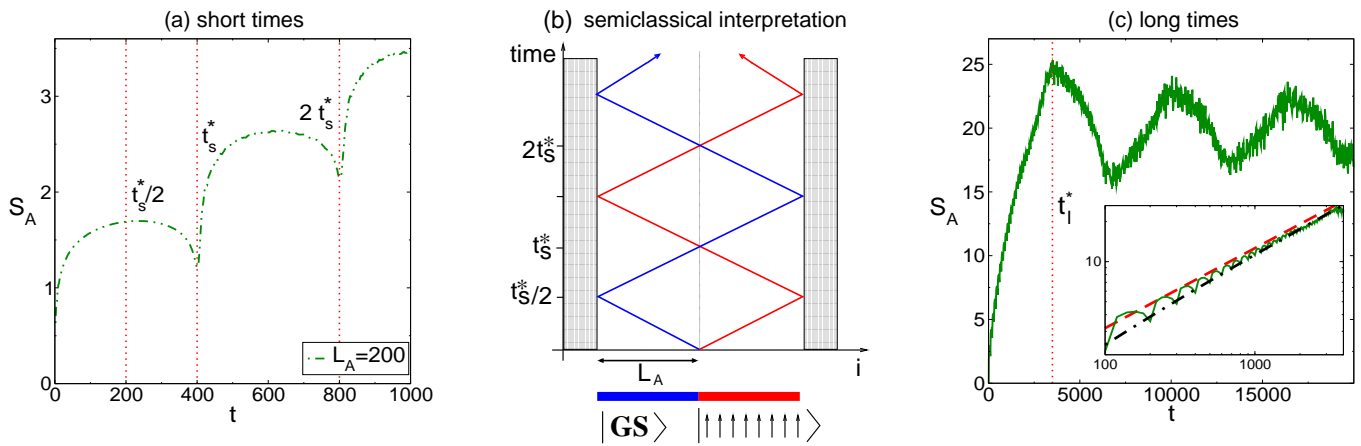}
\caption{ Entanglement spreading after the geometric quench with 
 aspect ratio $\omega\equiv L_A/L=1/2$ (cf. Fig.~\ref{fig00:quench_prot}) 
 in the open $XX$ chain. ({a}) von Neumann 
 entropy (dashed-dotted line) $S_A$ for part $A$ of the chain 
 as a function of time: exact results for $L_A=200$, $0\le t\le 
 t_s^*$ (short times), with $t_s^*\equiv 2L_A/v_s$ where $v_s=1$ is 
 the spinon velocity. Dotted vertical lines mark the times $t=t_s^*/2,
 t^*_s,2t_s^*$. ({b}) Semiclassical interpretation: entanglement 
 growth is understood in terms of the ballistic propagation 
 (with velocity $v_s=1$) of free effective excitations  
 (the lines denote  their trajectories). These are created at $t=0$ at the interface  
 between $A$ and $B$. At $t=t_s^*/2$,  perfect reflection at the boundary 
 of the chain occurs. Entanglement ``jumps'' at $t=m t^*_s,m=1,2\dots$ 
 correspond to excitations crossing the center of the  
 chain. ({c}) Long-time behavior: at $t\sim t_\ell^*\equiv L_A^2/v_s$ 
 (i.e., after ${\mathcal O}(L_A)$ crossings) the system reaches a steady state 
 with constant entropy (apart from superimposed oscillations). Inset: approach to the 
 steady-state entanglement (data for the same parameters as in the main figure and  
 $0\le t\le t_\ell^*$). A logarithmic scale is used on both axes. The 
 dashed line is $S_A\sim t^{\alpha}$, with $\alpha\approx 0.6$, whereas 
 the dashed-dotted one is $S_A\sim t^{1/2}\log(t)$.
}
\label{fig0:overview}
\end{figure*}

\section{Entanglement spreading in free systems: overview}
\label{overview}

We now turn to the real-time dynamics of the entanglement entropy 
$S_A$ between the two parts $A$ and $B$ of the chain. Here we consider 
the open $XX$ chain, restricting ourselves to an aspect ratio of 
$\omega=1/2$ (see Fig.~\ref{fig00:quench_prot}). The calculation of the 
entanglement entropy after the geometric quench in the $XX$ chain is 
outlined in Appendix~\ref{ent_free_geo}.

Clearly, at $t=0$, $A$ is in a pure state, implying $S_A=0$.
Exact numerical data at $t>0$ after the quench (with fixed $L_A=200$) 
are shown in Fig.~\ref{fig0:overview} [dashed-dotted line in panels ({a}) and 
({c})]. $S_A$ exhibits different behaviors at different time scales. At short 
times $t\le t_s^*/2$, with $t_s^*\equiv 2L_A/v_s$, (cf. Fig.~\ref{fig0:overview} 
({a})) the entanglement entropy grows logarithmically as in a local quench. 
In the time-interval $t_s^*/2\lesssim t\le t_s^*$, it slightly decreases 
reflecting the finite size of part $A$. 

At intermediate times $t_s^*<t\le t_\ell^*$ (cf. Fig.~\ref{fig0:overview} ({c})), 
$S_A$ grows with a power law (cf. the inset in Fig.~\ref{fig0:overview} ({c})). 
A fit to $S_A\sim t^{\alpha}$ yields $\alpha\approx 0.6$ (dashed line in the inset). 
However, we should stress that the data are also compatible with the behavior 
$S_A\sim t^{1/2}|\log(t)|$ (dashed-dotted line in the inset). A similar power-law 
increase of the entanglement entropy has been observed in quantum quenches in 
quasicrystals~\cite{igloi-2013}. Interestingly, partial revivals of the short-time 
dynamics are superposed with the power-law growth, in contrast with the local 
quench, where perfect revivals occur, apart from scaling 
corrections~\cite{stefan-2011}. 

The qualitative behavior of the entanglement can be understood in a semiclassical 
picture in terms of the ballistic propagation of the magnetization wavefront 
discussed in Sec.~\ref{wave}. This is illustrated in Fig.~\ref{fig0:overview} 
({b}). The initial entanglement increase at $t>0$ corresponds to the two edges of 
the wavefront (red and blue lines in the figure) propagating with equal velocities   
in the two parts of the chain. At $t=t^*_s/2$ the two edges are reflected at 
the physical boundaries. Finally, at $t=t_s^*$, a crossing of the 
two edge trajectories occurs. Every crossing at the later times $t=kt_s^*,\, k\in 
{\mathbb N}$, is reflected in an sudden increase in the von Neumann entropy 
(Fig.~\ref{fig0:overview} (a)). A similar semiclassical picture~\cite{calabrese-2006} 
holds in the case of a local quench~\cite{calabrese-2007a}, where the entanglement growth 
is associated with the propagation of two ``localized'' defects~\cite{eisler-2007,
eisler-2007a,peschel-2009}. 

At $t>t_\ell^*\sim L_A^2/v_s$, i.e., after ${\mathcal O}(L_A)$ crossings of the wavefront edge 
trajectories, (cf.~Fig.~\ref{fig0:overview} ({b})), the system reaches a steady state and 
the von Neumann entropy oscillates around a stationary value. We anticipate that, since the 
model is integrable \cite{rigol-2007}, the steady state is different from a thermal state, although its 
entanglement entropy is {\it extensive} (cf. Sec.~\ref{steady-state}).

\section{Short-time entanglement dynamics}
\label{short-time}

\begin{figure}[t]
\includegraphics[width=.95\linewidth]{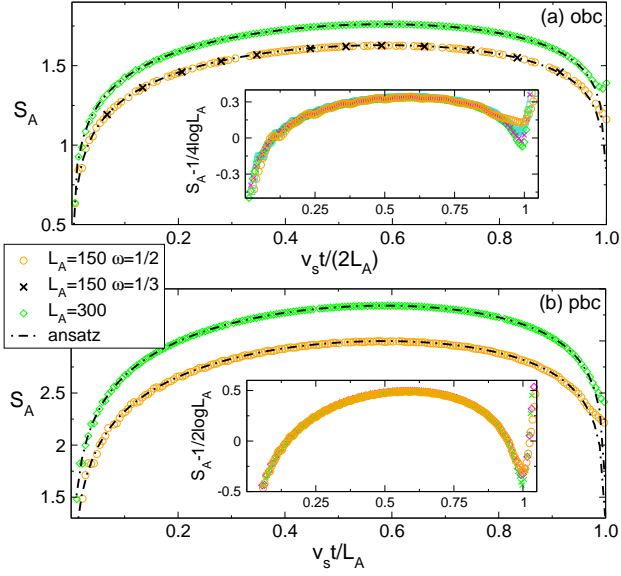}
\caption{von Neumann entropy $S_A$ for part $A$ after the 
 geometric quench (with $\omega\equiv L_A/L=1/2,1/3$) in the $XX$ chain: 
 short-time behavior (at $t\le t_s^*\sim  L_A/v_s$, and 
 $v_s=1$ the spinon velocity). ({a}) $XX$ chain with open boundary 
 conditions, $S_A$ versus $v_st/(2L_A)$ (exact results for several sizes 
 $L_A$). The crosses are data for $L=150$ and $\omega=1/3$. Dashed-dotted 
 lines are one parameter fits to $S_{\rm ansatz}$ (see Eq.~\eqref{short_ansatz}). 
 Inset: shifted entropy, $S_A-1/4\log L_A$ versus $v_st/(2L_A)$. Note 
 the perfect data collapse for all chain sizes and times. ({b}) The same 
 as in ({a}) for periodic boundary conditions: now $S_A$ is plotted versus 
 $v_st/L_A$. Dashed-dotted lines are fits to Eq.~\eqref{short_ansatz} 
 (notice the dependence on boundary conditions). Inset: shifted entropy, 
 $S_A-1/2\log L_A$  versus $v_st/L_A$.
}
\label{fig1:short}
\end{figure}

\begin{figure}[t]
\includegraphics[width=1\linewidth]{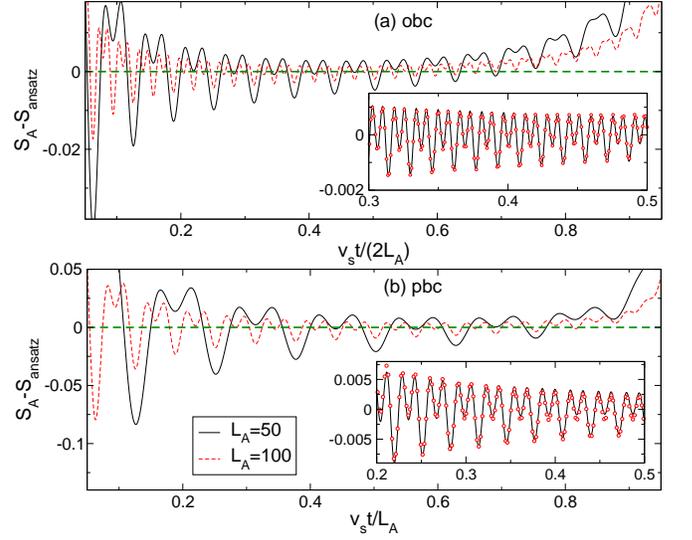}
\caption{Geometric quench ($\omega\equiv L_A/L=1/2$, cf. 
 Fig.~\ref{fig00:quench_prot}) in the $XX$ spin chain: 
 short-time dynamics of the von Neumann entropy 
 $S_A$ for part A of the chain. Deviations from $S_{\rm ansatz}$ 
 (see Eq.~\eqref{short_ansatz}) for ({a}) open and 
 (b) periodic boundary condition. (a) $S_A-S_{\rm ansatz}$ plotted versus 
 $v_st/(2L_A)$. Data are exact results for $L_A=50,100$ (full and dashed 
 line, respectively). Note that $k'_\nu$ (see Eq.~\eqref{short_ansatz}) 
 has been fitted and subtracted from the curves. Inset: same as in the 
 main panel (circles are data for $L_A=100$), 
 the continuous line is a fit to $[a_1\cos(t))+a_2\cos(2t)]/t$. 
 ({b}) Same as in ({a}) yet for periodic boundary conditions: 
 $S_A-S_{\rm ansatz}$ versus $v_st/L_A$. Inset: same as in ({a}). 
 The full line is a fit to $[a_1\cos(t)+a_2\cos(2t)]/t$. 
}
\label{fig2:corrections}
\end{figure}

In this section we focus on the short-time entanglement dynamics (i.e., at 
$t\le t_s^*$, cf. Fig.~\ref{fig0:overview} (a)). Here, in particular, we provide  
an analytic expression, which accurately describes the von Neumann entropy 
dynamics at short times $t\lesssim t_s^*\equiv\textrm{min}
(2L_A/(\nu v_s),2L_B/\nu)$. 
We motivate this formula based on heuristic arguments. This result holds 
irrespective of the quench aspect ratio $\omega$ 
(see Fig.~\ref{fig1:short} (a)), 
since the spreading of information between $A$ and $B$ is associated with the 
propagation of the two wavefront edges (cf. Fig.~\ref{fig0:overview} ({b}))
and part $B$ of the chain is prepared in the ``vacuum'' state. 

Figure~\ref{fig1:short} shows $S_A(t)$ as a function of the rescaled time 
$\nu v_st/(2L_A)\le 1$. Data are exact numerical results for the $XX$ chain 
with either open or periodic boundary conditions (panels ({a}) and 
({b}), respectively). For the sake of simplicity we restrict ourselves to a 
geometric quench with aspect ratio $\omega=1/2$. Motivated by the result 
for the local quench~\cite{calabrese-2007,igloi-2009,eisler-2008,stefan-2011} 
Eq.~\eqref{loc-cft} we have fitted the numerical data to 
\begin{equation}
S_{\rm ansatz}(t)=\alpha_\nu\log(t)+\beta_\nu\log\left[L_A\sin\frac{\nu\pi 
v_st}{2L_A}\right]+\gamma_\nu
\label{fit}
\end{equation}
where $\nu=1,2$ are for open and periodic boundary conditions, respectively, $v_s=1$ is 
the spinon velocity, and $\alpha_\nu,\beta_\nu,\gamma_\nu$ are fitting parameters. In Eq.~\eqref{fit} 
the first term is motivated by the fact that $S_A(t)$ is not symmetric under $t\to 2L_A/(\nu v_s)-t$, 
i.e., left-right inversion (see Fig.~\ref{fig1:short}), while the second one is similar to  
the local quench result Eq.~\eqref{loc-cft}. We have numerically found that $\beta_\nu=2
\alpha_\nu=\nu/6$. Finally, we rewrite Eq.~\eqref{fit} as 
\begin{equation}
S_{\rm ansatz}(t)=\frac{\nu}{6}\log\left|L_A^{\frac{3}{2}}\left(
\frac{\nu v_st}{2L_A}\right)^{\frac{1}{2}}\sin\frac{\nu\pi v_st}{2L_A}
\right|+k'_\nu, 
\label{short_ansatz}
\end{equation}
%
%
with $k'_\nu$ a constant. In the limit $t/L_A\ll 1$ (short times), one obtains 
from Eq.~\eqref{short_ansatz}  
\begin{equation}
S_A(t)=\frac{\nu}{4}\log t+k'_\nu
\label{short_asy_ansatz}, 
\end{equation}
which is different from the local-quench CFT result Eq.~\eqref{loc-asy}. 
A similar result is discussed in Ref.~\onlinecite{eisler-2014}, 
where the ground-state entanglement entropy of two free-fermionic chains 
connected by a narrow ``transition'' region  is studied. The two chains 
are completely full or empty, respectively, while a density variation in the 
transition region is induced by a linear chemical potential. Interestingly, 
the von Neumann entropy of a block that includes the transition region grows 
logarithmically with the block size, with a prefactor $1/4$, similar 
to Eq.~\eqref{short_asy_ansatz}.

It is useful to rewrite Eq.~\eqref{short_ansatz} as 
\begin{equation}
S_{\rm ansatz}=\frac{\nu}{6}\log\left|y^{\frac{1}{2}}\sin (\pi y)\right|
+\frac{\nu}{4}\log(L_A)+k'_\nu,
\label{short_ansatz_1}
\end{equation}
with $y$ being the rescaled time as in Eq.~\eqref{loc-cft}. Clearly, the 
shifted entanglement $S_A(t)-\nu/4\log(L_A)$ is a function of only $y$.

The validity of Eq.~\eqref{short_ansatz} is further corroborated by comparing with the data 
shown  in Fig.~\ref{fig1:short}: dashed lines in the two panels ({a}) and ({b}) are 
one-parameter fits to Eq.~\eqref{short_ansatz}, with $k'_\nu$ the only fitting parameter, 
which are in perfect agreement with the numerical results. In order to demonstrate the 
scaling behavior Eq.~\eqref{short_ansatz_1} we plot $S_A(t)-\nu/4\log L_A$ 
versus $\nu v_s t/(2L_A)$ in the insets of Fig.~\ref{fig1:short}. All data for different 
sizes collapse on the same curve, further confirming Eq.~\eqref{short_ansatz}. 

Finite-size deviations from Eq.~\eqref{short_ansatz}  are illustrated in 
Fig.~\ref{fig2:corrections}, plotting $S_A(t)-S_{\rm ansatz}(t)$ versus $\nu v_st/(2L_A)$ 
for $L_A=50,100$ (same data as in Fig.~\ref{fig1:short}). The constant $k'_\nu$ (cf. 
Eq.~\eqref{short_ansatz}) has been fitted and subtracted from the data. Finite-size 
corrections oscillate with time and vanish in the limit of large chains. We numerically 
checked that the formula 
\begin{equation}
S_A(t)-S_{\rm ansatz}(t)=\frac{1}{t}\big(a_1\cos(t)+a_2\cos(2t)\big)
\label{corr_ansatz}
\end{equation}
accurately describes the corrections at the intermediate time scales $0\ll t\ll 
L_A$, as shown in the inset in Fig.~\ref{fig2:corrections}. Symbols are data for $L_A=100$, 
while the continuous line is a fit to Eq.~\eqref{corr_ansatz}, with $a_1,a_2$ the 
fitting parameters. Notice the increasing behavior at $t\sim \nu L_A/v_s$, which could 
suggest a logarithmic correction as $\log(t)/t$. Similar corrections have been observed 
in the variance of the spin current after the quench from the ``domain wall'' state 
$|\cdots\uparrow\uparrow\uparrow\downarrow\downarrow\downarrow\cdots\rangle$~\cite{antal-2008}.

\section{Entanglement properties in the steady state}
\label{steady-state}

This section is devoted to studying entanglement properties in the steady 
state after a generic geometric quench with arbitrary aspect ratio $\omega$  
(cf. Fig.~\ref{fig00:quench_prot}). This corresponds to time scales $t\gg 
t_\ell^*$ (cf. Fig.~\ref{fig0:overview} ({c})). Here we restrict our 
analysis to the $XX$ chain with periodic boundary conditions. 

We first focus on the nature of the steady state after the quench. We show   
that it is {\it not} a thermal state,  
yet it reflects the initial half-filled Fermi sea in part $A$ of the 
chain. This observation allows us to derive an approximate analytic expression 
for the steady-state entanglement entropy, which is then checked 
against exact numerical results, finding good agreement, at least in the limit 
$L_A\ll L$. Remarkably,  despite the non-thermal nature of the state, its 
entanglement is {\it extensive}. This is also confirmed through direct inspection 
of the so-called single-particle entanglement spectrum (ES).

\begin{figure}
\includegraphics[width=.95\linewidth]{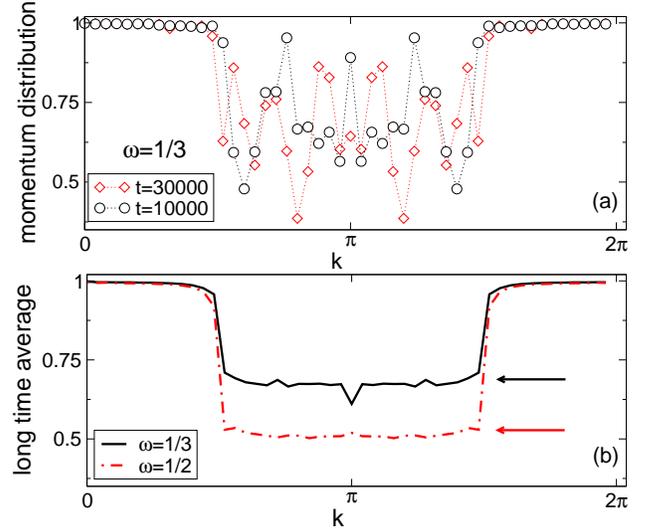}
\caption{ Momentum distribution function ${n}_k\equiv
 \langle c^\dagger_kc_k\rangle$ restricted to part $A$ of the chain 
 (cf. Fig.~\ref{fig00:quench_prot}) after a geometric quench in 
 the periodic $XX$ chain. Here $c_k,c_k^\dagger$ are 
 the fermionic operators of the corresponding free-fermion chain (cf. 
 Appendix~\ref{xx-chain}). ({a}) ${n}_k$ in the steady state 
 ($t\gg t_\ell^*$, cf. Fig.~\ref{fig0:overview}) plotted versus 
 the single-particle momentum $0\le k\le 2\pi$. Data are for $L_A=50$. 
 ({b}) Long-time average of ${n}_k$ (over the interval 
 $20000\le t\le 40000$). Full and dashed-dotted lines are  
 for the geometric quench with aspect ratios $\omega=1/3$ and 
 $\omega=1/2$, respectively (Fig.~\ref{fig00:quench_prot}). The plateaux 
 in the central region ($\pi/2\le k\le 3/2\pi$) correspond to ${n}_k=1-
 \omega$ (arrows in the figure). 
 }
\label{fig4:steady_mom}
\end{figure}

\subsection{Non-thermal steady state} 

\begin{figure}
\includegraphics[width=.95\linewidth]{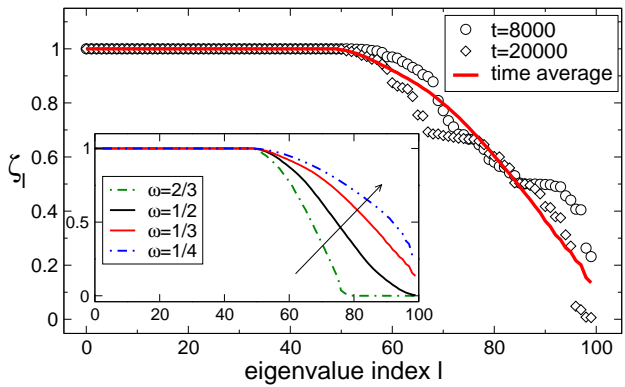}
\caption{Entanglement in the steady state after a geometric quench
 with $\omega=1/3$ in the periodic $XX$ chain. Single-particle 
 entanglement spectrum levels $\zeta_l$ ($l=1,2,\dots, 
 L_A$)  at long times $t>t_\ell^*$ (see Fig.~\ref{fig0:overview}). Data are 
 exact results for a chain with $L_A=100$ and several times $8000\le t\le 20000$. 
 Note  the first $L_A/2$ levels with $\zeta_l=1$. The full line is the long time 
 average. Inset: Long-time average of the single-particle entanglement 
 spectrum for quenches with several values of $\omega\equiv L_A/L=2/3,
 1/2,1/3,1/4$. The arrow indicates decreasing $\omega$. 
 Notice that the time averaged $\zeta_l$ are in general different from the 
 time averaged $n_k$ in Fig.~\ref{fig4:steady_mom} (b).
 }
\label{fig5:ES_asy}
\end{figure}

After a Jordan-Wigner transformation (cf. Appendix~\ref{xx-chain}) the 
$XX$ Hamiltonian, obtained from Eq.~\eqref{ham-xxz} imposing $\Delta=0$, is recast in 
a free-fermionic form as 
\begin{equation}
{\mathcal H}_{XX}=-\frac{1}{2}\sum\limits_{i=1}^{L-1}(c^\dagger_i 
c_{i+1}+c_ic^\dagger_{i+1})
\end{equation}
with $c_i$ standard fermionic operators. Entanglement properties in free-fermionic 
models (cf. Appendix~\ref{xx-chain}) are fully characterized by the two-point 
correlation function ${\mathbb G}_{m,n}\equiv\langle c_m^\dagger c_n\rangle$  
restricted to the subsystem, i.e., $m,n\in A$~\cite{peschel-1999,peschel-1999a,chung-2001,
peschel-2004,peschel-2004a,peschel-2009}. In the steady state, 
we numerically observe that the Fourier transform of $G_{m,n}$, $\widetilde{\mathbb G}_{k,k'}$ is 
approximately diagonal, i.e., $\widetilde{\mathbb G}_{k,k'}\approx n_k\delta_{k,k'}$, 
with ${n}_k$ the subsystem momentum distribution function  
\begin{equation}
\label{mom_dis}
n_k\equiv\langle c_k^\dagger c_k\rangle=\frac{1}{L_A}\sum\limits_{m,n}
e^{ik(m-n)}\langle c_m^\dagger c_n\rangle. 
\end{equation}
Here $k\equiv 2\pi s/L_A$, with $s=0,1,\dots L_A-1$, is the single-particle momentum,  
$\langle\cdot\rangle$ denotes the expectation value with the post-quench wavefunction 
at time $t$, and $m,n\in[1,L_A]$. 

$n_k$ is shown versus $k$ in Fig.~\ref{fig4:steady_mom} ({a}). Data are for the $XX$ chain 
with $L_A=50$ and the geometric quench with aspect ratio $\omega=1/3$. We restrict ourselves to 
times such that $t\gg t_\ell^*$ (cf. Fig.~\ref{fig0:overview}). Interestingly, apart from 
oscillations, a step-like structure is visible, which reflects the $t=0$ half-filled Fermi 
sea in part $A$ of the chain. 
The step-like form of $n_k$ is better visible in Fig.~\ref{fig4:steady_mom} ({b}). 
The continuous and dashed-dotted lines denote the time-averaged $n_k$ (in the 
interval $20000\le t\le 40000$) for two different quenches with $\omega=1/3$ 
and $\omega=1/2$. 

The form of $n_k$ can be derived in a semiclassical framework. 
At $t=0$ one can consider the excitations (particles) of the two independent chains $A$ and $B$ 
as uniformly distributed in each chain. Similarly, at $t\to\infty$ these are uniformly distributed  
in the final chain. Since the model is non-interacting, each mode preserves its momentum during 
the post-quench dynamics. The asymptotic (i.e., at $t\to\infty$) $n_k$ is then obtained as the ``average'' 
of the two initial Fermi seas $n_k^0(A)$ and $n_k^0(B)$ of parts $A$ and $B$, respectively. The 
result reads  
\begin{equation}
n_k=\omega\,n^0_k(A)+(1-\omega)\,n^0_k(B), 
\end{equation}
where $\omega=L_A/L$ and $1-\omega=L_B/L_A$ have to be interpreted as the probabilities that 
a mode (with given momentum) is occupied by a particle originally in $A$ and $B$, respectively. 
Notice that in the reasoning above we are considering $L,L_A\to\infty$, i.e., we neglect the 
finite lattice spacing. Using that $n_k^0(A)$ and $n_k^0(B)$ are the half-filled and the filled
Fermi seas, respectively, one obtains 
\begin{equation}
{n}_k=1-\omega\, 
\theta\left[k-\frac{\pi}{2}\right]\theta\left[\frac{3}{2}\pi-k
\right],
\label{mom_step}
\end{equation}
in perfect agreement with the numerical data in Figs.~\ref{fig4:steady_mom} (a) and (b). 
We should mention that similar results have been obtained studying the dynamics from the 
initial state $|m\rangle\otimes|-m\rangle$ in Ref.~\onlinecite{sabetta-2013}.

\subsection{Steady-state entanglement} 

The entanglement entropy of subsystem $A$ at any time can be given as (cf. 
Appendix~\ref{gs_ent_XX})
\begin{equation}
S_A(t)=-\sum\limits_l[\zeta_l\log\zeta_l+(1-\zeta_l)\log(1-\zeta_l)], 
\label{ent_peschel}
\end{equation}
where $\zeta_l$, which are related to the single-particle entanglement spectrum (ES) (cf. 
Appendix~\ref{gs_ent_XX}), are the eigenvalues of the two-point correlation matrix 
${\mathbb G}_{m,n}$ restricted to part $A$ of the chain.

\begin{figure}
\includegraphics[width=1\linewidth]{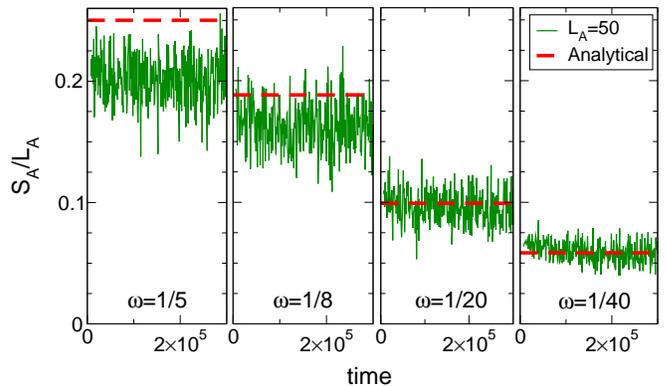}
\caption{ Extensive entanglement in the steady state after a geometric 
 quench. Data are for the periodic $XX$ chain. von Neumann 
 entropy $S_A$ for part $A$ of the chain (cf. Fig.~\ref{fig00:quench_prot}) 
 at large times ($t\gg t_\ell^*$, see Fig.~\ref{fig0:overview}): 
 $S_A/L_A$ plotted versus time $t$ for quenches with several aspect 
 ratios $\omega=L_A/L=1/5,1/8,1/20,1/40$ and $L_A=50$ (the same scale is 
 used on both axis in all the panels). The dashed line is $S_A/L_A=
 -[\omega\log\omega+(1-\omega)\log(1-\omega)]/2$.
}
\label{fig10:ent_asy}
\end{figure}

The behavior of $\zeta_l$ in the steady state is illustrated in Fig.~\ref{fig5:ES_asy}  
for the geometric quench with $\omega=L_A/L=1/3$ and fixed $L_A=100$. The continuous line 
is the (long) time average of the levels. At time $t>t_\ell^*$ there are $L_A/2$ (i.e., 
half of the levels) with $\zeta_l=1$, which do not contribute to the entropy (cf. 
Eq.~\eqref{ent_peschel}). The remaining $L_A/2$ are distributed over the whole interval 
$(0,1)$. The existence of an extensive number of levels with $\zeta_l\sim 1/2$ suggests 
that $S_A(t)$ is {\it extensive} in the steady state. This is dramatically different in the 
equilibrium ground state, where only few levels of the single-particle entanglement spectrum 
contribute in Eq.~\eqref{ent_peschel} (cf. Fig.~\ref{fig12:gs_sES} in Appendix~\ref{gs_ent_XX}).

The behavior of $\zeta_l$ upon varying the aspect ratio 
$\omega$ is illustrated in the inset in Fig.~\ref{fig5:ES_asy}, showing 
the time-averaged levels for $\omega=2/3,1/2,1/3,1/4$. Irrespective of 
$\omega$  an extensive fraction ($\sim L_A/2$) of levels is in the region 
$\zeta_l\approx 1$. Moreover, as $\omega$ decreases the whole distribution is 
shifted towards $\zeta_l=1$,  signaling that, although extensive behavior 
persists at any $\omega$, the actual value of the entropy decreases as 
$\omega\to 0$. 

Finally, the scenario outlined above can be justified using 
Eq.~\eqref{mom_step}, i.e., neglecting the oscillations in 
Fig.~\ref{fig4:steady_mom} ({a}). Within this approximation, 
${\mathbb G}_{m,n}$ has $L_A/2$ {\it identical} eigenvalues $\zeta_l=1-\omega$ (and $L_A/2$ 
unit eigenvalues). Notice that these are different from 
the long time average of $\zeta_l$ in Fig.~\ref{fig5:ES_asy}, suggesting 
that the diagonal approximation $\widetilde{\mathbb G}_{k,k'}
\approx n_{k}\delta_{k,k'}$ might be too crude.  
The von Neumann entropy, using Eq.~\eqref{ent_peschel} 
is then 
\begin{equation}
\frac{S_A}{L_A}=-\frac{1}{2}\left[\omega\log\omega+(1-\omega)
\log(1-\omega)\right]. 
\label{asy_ent}
\end{equation}
The comparison between Eq.~\eqref{asy_ent} and the exact data is shown in Fig.~\ref{fig10:ent_asy}, 
focusing on the geometric quenches with $\omega=1/5,1/8,1/20,1/40$ (panels from left to 
right in the figure). The continuous lines are data for $S_A/L_A$ at fixed $L_A=50$ and  
$10^4<t<10^6$. Clearly, $S_A/L_A\to0$ upon decreasing $\omega$, as expected (cf. 
inset in Fig.~\ref{fig5:ES_asy}). In each panel, the value according to Eq.~\eqref{asy_ent} is 
shown as a dashed line. At $\omega=1/5$ and $\omega=1/8$ some deviations from Eq.~\eqref{asy_ent} 
are observed, which have to be interpreted as finite-size effects, due to the fact that $L_A\sim L$. 
In the limit $L_A\ll L$ (equivalent to $\omega\to 0$), Eq.~\eqref{asy_ent} is in remarkably good 
agreement with the exact data.

\begin{figure}
\includegraphics[width=.97\linewidth]{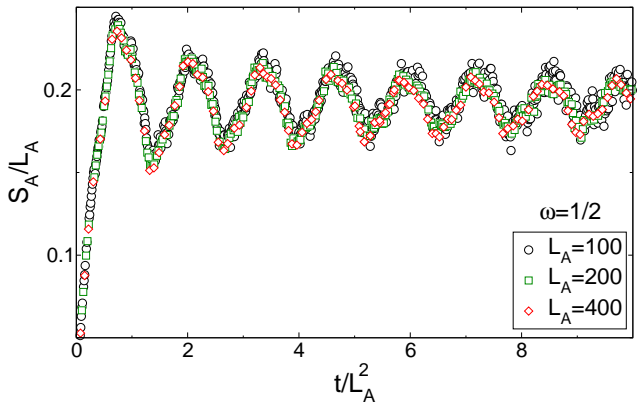}
\caption{ Fluctuations of the entanglement entropy in the steady  
 state after a geometric quench with aspect ratio $\omega\equiv 
 L_A/L=1/2$. Data are for the open $XX$ chain with $L=100,200,400$: 
 rescaled von Neumann entropy $S_A/L_A$ plotted versus the rescaled 
 time $t/L_A^2$. Notice that partial revivals (oscillations) persist 
 in the long time regime, and do not decay with the chain size. 
}
\label{fig8a:steady_fluct}
\end{figure}

It is interesting to investigate the entanglement fluctuations in the 
steady state. These are illustrated in Fig.~\ref{fig8a:steady_fluct} plotting the 
rescaled von Neumann entropy $S_A/L_A$ versus $t/L_A^2$. Data are for the 
open $XX$ chain with $L=100,200,400$ and a geometric quench with aspect ratio 
$\omega=1/2$ (cf. Fig.~\ref{fig00:quench_prot}). Here we focus on intermediate and 
long time scales after the quench, i.e., $t\gg t_s^*$ (see Fig.~\ref{fig0:overview}). 
Remarkably, in Fig.~\ref{fig8a:steady_fluct} all the data collapse on the same curve,    
confirming that the steady state exhibits extensive entanglement, in agreeement with 
the semiclassical result Eq.~\eqref{asy_ent}. However, oscillating deviations from the 
steady state value are observed, with period $\sim 1/L_A^2$ and amplitude $\sim L_A$. 
Notice that at fixed $t/L_A^2$ these oscillations do not vanish in the limit $L_A\to\infty$.

\section{Entanglement relaxation after the infinite chain quench}
\label{inf_size_quench}

In this section we discuss the real-time entanglement dynamics after the infinite-chain 
geometric quench, which corresponds to the limit $\omega\to 0$, at fixed {\it finite} 
$L_A$ (see Fig.~\ref{fig00:quench_prot}). Here we focus on time scales $t>t_s^*$ 
(for $t<t_s^*$, one has the same behavior as in Sec.~\ref{short-time}), considering 
the $XX$ chain with periodic boundary conditions. Although we are interested in 
the limit $\omega\to 0$, in practice we consider finite (large) $\omega$ restricting 
ourselves only to $t<L-L_A=L_A(\omega^{-1}-1)$, to avoid reflections at the boundaries 
of the chain.

First, one has $S_A(t)\to 0$ at $t\to\infty$, since the wavefunction becomes a product state 
in the limit $\omega\to 0$. However, the entanglement relaxation dynamics, at any time $t>t_s^*$, 
is described by a scaling function $f_\ell(z)$ of the rescaled time $z=t/L_A^2$. Additionally, 
two different dynamical regimes appear: while $f_\ell(z)\sim -\log(z)$ at $z\lesssim 1$, a 
crossover to $f_\ell(z)\sim 1/z+1/z\log(z)$ occurs around $z\sim 1$. The latter behavior can 
be calculated analytically.

\begin{figure}
\includegraphics[width=1\linewidth]{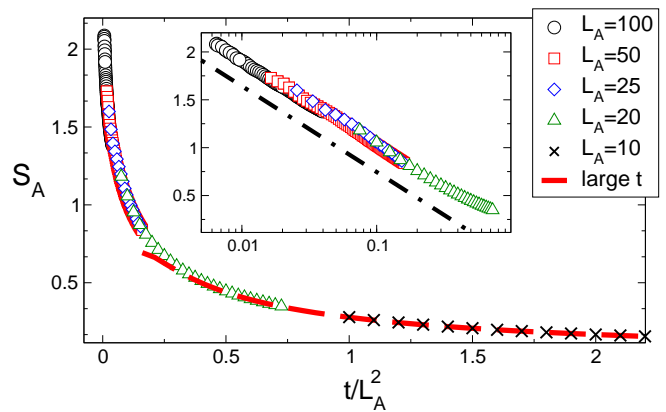}
\caption{ Entanglement relaxation after the geometric quench with   
 aspect ratio $\omega\equiv L_A/L\ll 1$ in the periodic $XX$ chain. 
 Symbols are exact numerical data for $L_A=10,20,25,50,100$. 
 Rescaled entropy dynamics: $S_A$ versus $t/L_A^2$. Perfect data 
 collapse is observed for all $L_A$ and times. The dashed line 
 is the analytic result at $t\to\infty$. Inset: same as in the 
 main figure at $t/L_A^2<1$. Note the logarithmic scale on the 
 $x$-axis. The dashed dotted line is $-0.4\log(t/L_A^2)$.
 }
\label{fig6:relaxation}
\end{figure}

All these features are present in the data shown in Fig.~\ref{fig6:relaxation}, plotting 
$S_A(t)$ versus $z\equiv t/L_A^2$ for a geometric quench with $\omega=1/30$ and several $L_A$. 
The perfect data collapse provides robust evidence that the entanglement dynamics at $t>t_s^*$ 
is described by a scaling function $f_\ell(z)$. The behavior of $f_\ell(z)$ at $z\to\infty$ is 
given analytically as (cf. Appendix~\ref{large-t})
\begin{equation}
S_A(t)\approx\frac{L_A^2}{2\pi t}\left[1-\log 
\frac{L_A^2}{2\pi t}\right]. 
\label{large-t_ent}
\end{equation}
This is shown in Fig.~\ref{fig6:relaxation} as a dashed line, in perfect 
agreement with the numerical data already at $t/L_A^2\sim 1/2$. On the other 
hand, at $t/L_A^2\ll 1$, one has the strikingly different behavior $S_A(t)= 
-\alpha\log(t/L_A^2)+\beta$, as numerically demonstrated in the inset of 
Fig.~\ref{fig6:relaxation}. In particular, a fit of the numerical data gives 
$\alpha\approx 0.4$ (dashed-dotted line in the figure).

\section{Geometric quench in the $XXZ$ chain}
\label{gq_xxz}

We now turn to the post-quench dynamics in interacting models, considering the 
$XXZ$ chain in the gapless phase at $-1<\Delta\le 1$. We restrict ourselves 
to short time scales, which can be accessed efficiently using 
tDMRG~\cite{daley-2004,feiguin-2004,schollwoeck-2011,schollwoeck-2005}.   

We provide numerical evidence that qualitative and quantitative features 
are similar to the $XX$ chain. First, after the quench a magnetization 
wavefront forms, spreading ballistically in the two parts of the chain 
$A$ and $B$ (cf. Fig.~\ref{fig00:quench_prot}). However, while in the $XX$ 
chain the two wavefront edges propagate with the same velocity (i.e., 
$v=1$, see Sec.~\ref{wave}), here two different velocities appear. More 
precisely, the propagation in parts $A$ and $B$ happens at the spinon velocity 
$v_s(\Delta)$ and $v=1$, respectively. Interestingly, we find that the central 
region of the wavefront is described by Eq.~\eqref{wf_app}. 

On the other hand, the entanglement dynamics is well described by the same 
formula derived for the $XX$ chain (cf. Eq.~\eqref{short_ansatz}). However, as 
the wavefronts (and consequently the information) propagation is  
anisotropic  in the two parts of the chain, a remarkable difference is that one 
has to replace $v_s\to v_e$ in Eq.~\eqref{short_ansatz}, with $v_e$ being an 
effective entanglement spreading rate. We numerically find that $v_e\approx v_s$ 
for $\Delta<0$ (i.e., the entanglement spreads with the wavefront edge velocity), 
whereas one has $v_e<v_s$ at $\Delta>0$.

\begin{figure}
\includegraphics[width=1\linewidth]{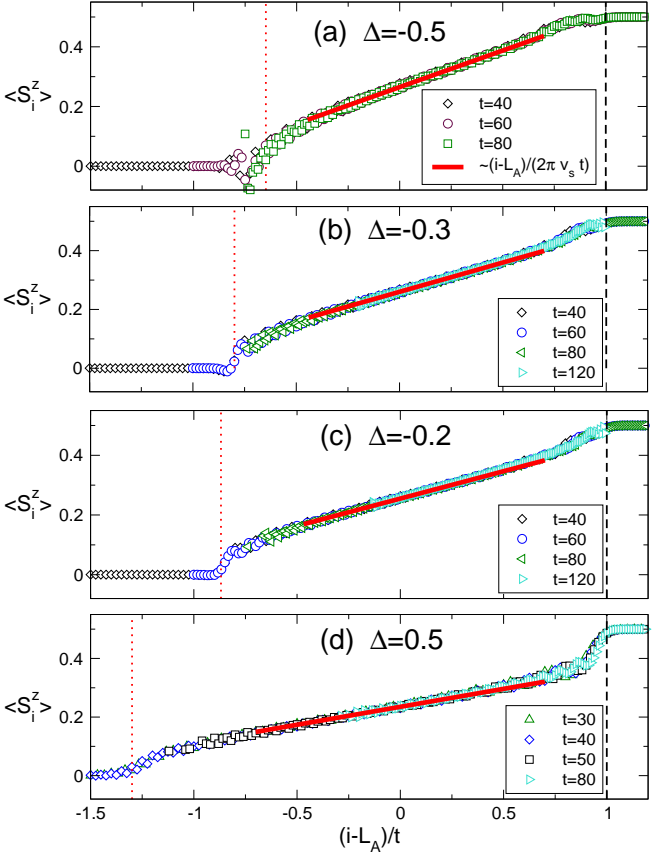}
\caption{ Magnetization wavefront after a geometric quench with 
 aspect ratio $\omega\equiv L_A/L=1/3$ in the  $XXZ$ spin chain. 
 Symbols are tDMRG data for $L_A=60$ and anisotropies  
 $\Delta=-0.5,-0.3,-0.2,0.5$ (panels (a)-(d) in the figure). The same 
 scale is used on the $x$-axis in all panels. The local magnetization 
 $\langle S_i^z\rangle$ is plotted 
 versus $(i-L_A)/t$, with $i$ being the position in the chain. 
 All the data collapse on the same $\Delta$ dependent scaling 
 function. The dashed vertical line corresponds to $(i-L_A)/t=1$, 
 while the dashed-dotted one is $(i-L_A)/t=-v_s(\Delta)$, with 
 $v_s(\Delta)$ the spinon velocity. The continuous 
 lines are  fits to $s_0+(i-L_A)/(2\pi v_s(\Delta) t)$, with 
 $s_0\approx 1/4$ the fitting parameter.
}
\label{fig7:wavefront_int}
\end{figure}

\subsection{Ballistic wavefront propagation}
\label{wave_int}

The local magnetization $\langle S^z_i(t)\rangle$ as a function of the 
site position $i$ in the chain is shown in Fig.~\ref{fig7:wavefront_int} 
for several times after  the geometric quench.  We restrict 
ourselves to $\omega=1/3$, showing data at fixed $L_A=60$ and several 
values of the anisotropy $\Delta$ in the spin-liquid phase ($\Delta=-0.5,
-0.3,-0.2,0.5$, panels (a)-(d) in the figure). Symbols denote 
tDMRG data for an $XXZ$ chain with open boundary conditions and  
$t\ll L_A$ to avoid effects from reflections at the boundary of the 
chain.

The formation of two propagating wavefronts at $t>0$ is clearly visible 
for all values of $\Delta$. Their left and right edges propagate 
ballistically in the two parts of the chain (see also Ref.~\onlinecite{langer-2009,langer-thesis}). 
This is illustrated plotting $\langle S_i^z\rangle$ versus the rescaled variable 
$(i-L_A)/t$. At each $\Delta$, the data collapse on the same function 
for all times and positions. The vertical dashed-dotted line in 
Fig.~\ref{fig7:wavefront_int} marks the point $(i-L_A)/t=-v_s$.  For 
the $XXZ$ chain in the zero magnetization sector $v_s$ is given as~\cite{gaudin-1966} 
\begin{equation}
\label{vf_ba}
v_{s}(\Delta)=\frac{\pi}{2}\frac{\sin\gamma}{\gamma}\quad 
\mbox{with}\quad \cos\gamma=\Delta\,.
\end{equation}
Clearly, $\langle S_i^z\rangle=0$ at $(i-L_A)/t=-v_s$, demonstrating  that the 
left edge of the wavefront propagates with $v_s$. 
On the other hand, the right one propagates at unit velocity (the dashed lines 
in all panels mark the point $(i-L_A)/t=1$). Although the full 
scaling function $\langle S_i^z\rangle\equiv m((i-L_A)/t)$ is not easily 
accessible, at $(i-L_A)\ll t$ (central region in the panels in 
Fig.~\ref{fig7:wavefront_int}) the magnetization profile exhibits the linear 
behavior $\langle S_i^z\rangle\sim (i-L_A)/t$. It is reasonable that this is 
given analytically by (as a generalization of Eq.~\eqref{wf_app})  
\begin{equation}
\langle S_i^z\rangle\approx s_0+\frac{i-L_A}{2\pi v_s(\Delta) 
t}\,.
\label{wf_int_app}
\end{equation}
The validity of Eq.~\eqref{wf_int_app} is confirmed in Fig.~\ref{fig7:wavefront_int}. 
Continuous lines in the figure are fits to Eq.~\eqref{wf_int_app} (with $s_0\approx 1/4$ 
the fitting parameter), and are in excellent agreement with the tDMRG data.

\subsection{Short-time entanglement dynamics}
\label{ent_dyn_int}

\begin{figure}
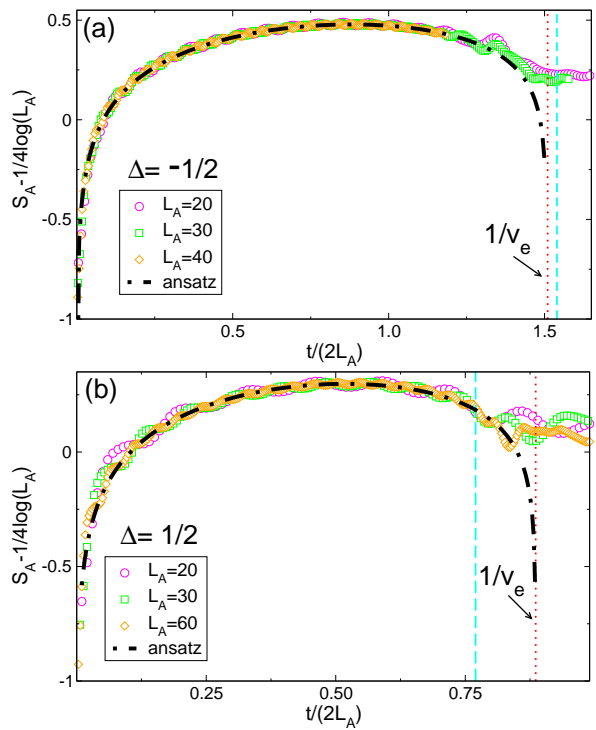

\includegraphics[width=.9\linewidth]{./fig11_stxxz}
\includegraphics[width=.9\linewidth]{./fig11_stxxz_1}
\caption{ Entanglement spreading after the geometric 
 quench with aspect ratio $\omega\equiv L_A/L=1/3$ (see 
 Fig.~\ref{fig00:quench_prot}) in the $XXZ$ spin chain at 
 $\Delta=-1/2$ ({a}) and $\Delta=1/2$ ({b}). ({a}) 
 Shifted von Neumann entropy $S_A-1/4\log L_A$ versus 
 $t/(2L_A)$. Symbols are tDMRG data for $L_A=20,30,40$. 
 Notice the perfect data collapse for all sizes and times. 
 The dashed line are fits to $S_{\rm ansatz}$ 
 (Eq.~\eqref{short_ans_int}), with $v_e$ (entanglement 
 spreading rate) and $k'_\nu$ the fitting parameters. The vertical 
 dotted line marks the point $2L_A/t=v_e$. The fit gives 
 $v_e\approx v_s(\Delta)$. ({b}) Same as in ({a}) for $\Delta=1/2$ 
 and $L_A=20,30,60$. The fit to Eq.~\eqref{short_ans_int} now 
 yields $v_e\approx 1.13<v_s\approx 1.3$.
 In both panels, the vertical dashed line is $1/v_s(\Delta)$. }
\label{fig9:short_int}
\end{figure}

In this section we investigate the entanglement spreading after a 
geometric quench in the (open) $XXZ$ chain in the gapless phase (i.e., 
$-1<\Delta\le 1$). At short time scales that can be accessed by tDMRG 
it is natural to generalize the result at $\Delta=0$ (cf. 
Eq.~\eqref{short_ansatz}) as 
\begin{equation}
S_{\rm ansatz}(t)=\frac{\nu}{6}\log\left[L_A^{\frac{3}{2}}\left(
\frac{\nu\, v_et}{2L_A}\right)^{\frac{1}{2}}\sin\frac{\nu\pi v_et}
{2L_A}\right]+k'_\nu
\label{short_ans_int}
\end{equation}
with $\nu=1$ for open and $\nu=2$ for periodic boundary conditions, 
$k'_\nu$ a $\Delta$-dependent constant, and $v_e$ an entanglement 
spreading rate. Equation~\eqref{short_ans_int} is expected to hold 
at $t\ll 2L_B$, although we are not able to provide its precise regime of 
validity, which would require the exact expression for $v_e$. 
From Eq.~\eqref{short_ans_int} one finds that $S_A(t)-\nu/4
\log (L_A)$ is a scaling function of $t/L_A$. The validity of Eq.~\eqref{short_ans_int} 
is shown in Fig.~\ref{fig9:short_int}, considering tDMRG data for the $XXZ$ chain at 
$\Delta=-1/2$ (panel ({a}) in the figure) and $\Delta=1/2$ (panel ({b})). We provide 
data for  $L_A=20,30,40,60$, restricting ourselves to a geometric quench with 
$\omega=1/3$. Strikingly, all the data for different system sizes collapse on the same scaling 
curve, in agreement with Eq.~\eqref{short_ans_int}. To further 
proceed we fit the data to Eq.~\eqref{short_ans_int} ($k'_\nu$ and $v_e$ being 
the only fitting parameters). Remarkably, at $\Delta=-1/2$, we  obtain 
$v_e\approx v_s$ (the vertical dotted line in Fig.~\ref{fig9:short_int} 
marks the point at $2L_A/t=v_s$). Also, we numerically verified that 
 $v_e\approx v_s$ in the whole interval $-1<\Delta\le 0$. 
However,  at $\Delta=1/2$ we obtain $v_e\approx 1.13<v_s\approx 1.3$ 
(the vertical-dashed line in Fig.~\ref{fig9:short_int} marks the point 
$t/(2L_A)=1/v_s$). Our analysis suggests that although the information 
spreading between the two parts $A$ and $B$ of the chain is associated 
with the wavefront propagation, the spreading rate $v_e$ is not a trivial  
function of the wavefront edges' velocities.

\section{Summary and conclusions} 
\label{summary}

In this work we investigated the entanglement dynamics after a geometric 
quench in the $XXZ$ chain in the gapless phase, both analytically and numerically. 
The initial state after the quench is obtained joining two chains $A$ and $B$, of 
lengths $L_A$ and $L_B$, prepared in the ground state of the $XXZ$ 
chain in the sector with zero and maximum magnetization, respectively. 
The latter is the fully polarized state, which can be a high-energy eigenstate 
of the model, depending on the exchange anisotropy. Equivalently, in the language 
of interacting fermions confined in a 
hard-wall trap, the geometric quench  corresponds to a sudden change in the 
trap size. From the energy point of view, this quench falls into the class 
of global quenches, since the excess energy density above the ground state 
of the final Hamiltonian is finite. On the other hand, both $A$ and $B$ are 
in eigenstates of the $XXZ$ model, implying that the post-quench dynamics 
originates locally from a ``defect'' at the interface between the two chains, 
as in typical local quenches. 

The entanglement growth after the quench is associated with the formation of 
a magnetization wavefront, whose edges propagate {\it ballistically} in the 
two parts of the chain, at different velocities. To be precise, while the wavefront 
expands in part $A$ with the spinon velocity, in part $B$ this happens at unit 
velocity. For the $XX$  chain we derived the exact analytical expression of the 
wavefront profile using free-fermionic techniques and semiclassical arguments. 
For the $XXZ$ model we found that the central region of the wavefront is 
described by a simple function, which depends on the spinon velocity.

Focusing on the $XX$ chain we observed that the entanglement dynamics 
after the quench exhibits several interesting dynamical regimes. Specifically, 
at $t\le t_s^*\sim L_A/v_s$ (short time scales) the von Neumann entropy increases 
logarithmically. Moreover, while the well-known CFT result~\cite{calabrese-2007,
calabrese-2007a,stefan-2011} for the entanglement growth in local quenches does 
not apply, we provided an analytic formula, derived from heuristic 
arguments, that  describes accurately the short-time entanglement dynamics. 
Remarkably, the entanglement entropy exhibits the scaling  behavior $S_A(t)=f_s(y)+
s(L_A)$, with $y=t/L_A$, $s(L_A)=\nu/4\log(L_A)$ (here $\nu=1,2$ for open and periodic 
boundary conditions, respectively), and $f_s(y)$ a scaling function. 
At larger times ($t_s^*\le t\le t^*_\ell$, with $t^*_\ell\sim L_A^2/v_s$) 
the von Neumann entropy shows partial revivals of the short-time 
dynamics superposed with a power-law increase $S_A\sim t^\alpha$. We 
numerically found $\alpha<1$. 

At very long times $t\ge t^*_\ell$ the system reaches a steady state 
and the entanglement entropy saturates, apart from oscillations. As expected, 
since the model is integrable, the steady state is {\it not} thermal. More 
precisely, we observed that the subsystem momentum distribution function 
shows discontinuities at $k=\pi/2$, reflecting the initial half-filled Fermi 
sea in part $A$ of the chain. Finally, we provided numerical and 
analytical evidence that the steady-state entanglement is {\it extensive}. 

We also considered the geometric quench from a finite to the infinite chain. 
While at large times one has $S_A\to 0$, reflecting the wavefunction being 
an almost perfect product state, the entanglement relaxation dynamics 
exhibits the scaling form $S_A(t)=f_\ell(z)$, with $z\equiv t/L_A^2$. 
Interestingly,  the behavior of $f_\ell(z)$ changes dramatically at $z\sim 1$ 
(i.e., $t\sim L_A^2$). Namely, we numerically observed that $f_\ell(z)\approx-
\gamma\log(z)$, with $\gamma\approx 0.4$ at $z\lesssim 1$. On the other hand, 
at $z\gtrsim 1$ we derived analytically $f_\ell(z)\approx 1/z(1-\log(1/z))$.

Finally, by means of tDMRG simulations we discussed the role of interactions in 
the short-time entanglement dynamics, considering the $XXZ$ chain in the gapless 
phase $-1<\Delta\le 1$. Interestingly, we numerically demonstrated that the 
same formula conjectured for the free-fermion case fully reproduces the 
short-time entanglement dynamics after the quench. Due to the 
anisotropic propagation of the wavefront in the two parts of the chain $A$ 
and $B$, the entanglement spreading rate is not trivially related to the 
velocity of a single wavefront edge.

\section{Acknowledgements}

We acknowledge very fruitful discussions with J.-S.~Caux, P.~Calabrese, 
M.~Fagotti, M.~Haque, J.~Mossel, G.~Palacios, P.~Ribeiro. V.A. 
thanks J.-S.~Caux, J.~Mossel, and G.~Palacios for valuable discussions 
in a related project, from which this work originated. We thank 
F.~Igl\'oi and J.~H.~H.~Perk for useful comments on a previous version 
of the manuscript. 

\appendix 
\section{Diagonalization of the spin-$1/2$ $XX$ chain}
\label{xx-chain}

The spin-$1/2$ open $XX$ chain~\cite{lieb-1961,barouch-1970,barouch-1971,barouch-1971a,mccoy-1971} 
of length $L$ in an external magnetic field $h$ is defined as  
\begin{equation}
{\mathcal H}_{XX}=-J\sum\limits_{i=1}^{L-1}(S^x_iS^x_{i+1}+S^y_i
S^y_{i+1})+h\sum\limits_{i=1}^{L}S_i^z, 
\label{xx_ham}
\end{equation}
with $S^{x,y,z}_i\equiv\sigma_i^{x,y,z}/2$, $\sigma_i^\alpha$ 
being the Pauli matrices acting on site $i$. For periodic 
boundary conditions one has an extra term in Eq.~\eqref{xx_ham} 
connecting site $L$ with site $1$. Hereafter we fix $J=1$ for 
convenience. After the Jordan-Wigner transformation 
\begin{equation}
c_i=\Big(\prod\limits_{m=1}^{i-1}\sigma^z_m\Big)
\frac{\sigma_i^x-i\sigma_i^y}{2},
\label{j-wigner}
\end{equation}
Eq.~\eqref{xx_ham} is recast in the free-fermionic form 
\begin{equation}
{\mathcal H}_{XX}=-\frac{1}{2}\sum\limits_{i=1}^{L-1}(c^\dagger_i 
c_{i+1}+c_ic^\dagger_{i+1})+\frac{h}{2}\sum\limits_{i=1}^{L-1}
c^\dagger_i c_i,
\label{xx_fer}
\end{equation}
with $c_i$ spinless fermionic operators satisfying the canonical 
anticommutation relations $\{c_m,c^\dagger_n\}=\delta_{m,n}$. 
Notice in Eq.~\eqref{j-wigner} the non local term (Jordan-Wigner string) 
in the brackets. The mapping between Eq.~\eqref{xx_ham} and Eq.~\eqref{xx_fer} 
is exact apart from boundary terms (that we neglect here) giving a 
vanishing contribution (as $1/L$) to physical quantities in the  large chain 
limit.

\paragraph*{Periodic boundary conditions (pbc).---} For periodic 
boundary conditions the spectrum of Eq.~\eqref{xx_fer} can be obtained 
going to momentum space. After defining the Fourier transformed 
operators $c_k$ as 
\begin{equation}
c_k=\frac{1}{\sqrt{L}}\sum_{m=1}^{L} e^{i\frac{2\pi k}{L} m}c_m, 
\end{equation}
and substituting in Eq.~\eqref{xx_fer}, one obtains the single-particle 
dispersion $E_k$ of the $XX$ chain as 
\begin{equation}
E_k=-\cos\frac{2\pi k}{L}+h\quad \mbox{with}\quad k=0,1,\dots, L-1, 
\label{disp_per}
\end{equation}
with $2\pi k/L$ and $E_k$  the single-particle 
momenta and energies, respectively. The single-particle eigenstate $|v_k\rangle$ 
corresponding to the eigenvalue $E_k$ reads  
\begin{equation}
|v_k\rangle=\frac{1}{\sqrt{L}}\sum\limits_{m=1}^{L}
e^{i\frac{2\pi k}{L} m}c_m^\dagger|0\rangle, 
\label{per_eig}
\end{equation}
with $|0\rangle$ denoting the vacuum state for the fermions.

\paragraph*{Open boundary conditions (obc).---} For the $XX$ chain 
with open boundary conditions one has instead   
\begin{equation}
    E'_k=-\cos\frac{\pi k}{L+1}+h\quad \mbox{with}\quad k=0,1,\dots, L-1
\label{disp_open},
\end{equation}
while the single-particle eigenstates are 
\begin{equation}
|v'_k\rangle=\sqrt{\frac{2}{L+1}}\sum\limits
\limits_{m=1}^L\sin\left[\frac{\pi mk}{L+1}\right] 
c^\dagger_m|0\rangle.
\label{op_eig}
\end{equation}
The spectrum  of the model  (cf. Eq.~\eqref{disp_per} and  Eq.~\eqref{disp_open}) 
is gapless in the thermodynamic limit at $|h|<1$, while it is gapped otherwise. 
The ground state at $h=1$ ($h=-1$) corresponds to an empty (fully filled) band. 

The ground state of the $XX$ chain is obtained by filling the single-particle 
levels (cf. Eqs.~\eqref{disp_per} and \eqref{disp_open}) below the Fermi level 
$k_F=L/(2\pi)\cos^{-1}(h)$ and $k_F=(L+1)/\pi\cos^{-1}(h)$ for periodic and open 
boundary conditions, respectively. Notice that for convenience, the Fermi level 
$k_F$ is defined as an integer. In this work we restrict ourselves to the 
$XX$ chain with zero magnetic field ($h=0$).

\section{Entanglement entropies in free-fermionic chains}
\label{gs_ent_XX}

Here we briefly review how to calculate the entanglement entropy for a 
generic eigenstate of a free-fermionic model~\cite{peschel-1999,peschel-1999a,
chung-2001,peschel-2004,peschel-2004a,peschel-2009,latorre-2009}, focusing, in particular, 
on the ground-state entropy. The von Neumann entropy (and Renyi entropies as 
well) of a {\it single} interval $A\equiv [1,L_A]$ (of length $L_A$) embedded in 
a free-fermionic chain can be obtained from the two-point correlation function 
${\mathbb G}_{m,n}$ restricted to the subsystem $A$ 
\begin{equation}
{\mathbb G}_{m,n}=\langle c^\dagger_m c_n\rangle\qquad \mbox{with}\quad m,n=1,2,\dots L_A.
\label{c2}
\end{equation}
Here $\langle\cdot\rangle$ denotes the expectation value over 
a generic eigenstate of Eq.~\eqref{xx_fer}. 

\paragraph*{Ground-state correlation matrix.---} The correlation matrix 
${\mathbb G}_{m,n}$ (cf. Eq.~\eqref{c2}) for a generic eigenstate of 
Eq.~\eqref{xx_fer} can be obtained using the explicit form of the 
single-particle eigenvectors Eq.~\eqref{per_eig} and Eq.~\eqref{op_eig}. For 
the ground state of Eq.~\eqref{xx_fer} and periodic boundary 
conditions one obtains 
\begin{equation}
{\mathbb G}^{({\rm pbc})}_{m,n}=\frac{2}{L}\sum\limits_{k=0}^{L/4-1}
\cos\left[\frac{2\pi (m-n)k}{L}\right]-\frac{1}{L}.
\label{c2_pbc}
\end{equation}
Performing the summation over $k$ one obtains
\begin{equation}
{\mathbb G}^{({\rm pbc})}_{m,n}=\frac{1}{L}\frac{\sin\left[(\frac{\pi}{2}-
\frac{\pi}{L})(m-n)\right]}{\sin\left[\frac{\pi}{L}(m-n)\right]}.
\label{c2_pbc_sum}
\end{equation}
In the limit of an infinite chain, Eq.~\eqref{c2_pbc_sum} reduces to 
\begin{equation}
{\mathbb G}^{({\rm pbc})}_{m,n}(L\gg 1)\to\frac{\sin[\frac{\pi}{2}(m-n)]}
{\pi(m-n)}.
\end{equation}
Finally, for open boundary conditions ${\mathbb 
G}_{m,n}^{({\rm obc})}$ reads 
\begin{multline}
{\mathbb G}^{({\rm obc})}_{m,n}=\frac{1}{2(L+1)}\Big[\frac{\sin \frac{\pi}{2}
(m-n)}{\sin \frac{\pi}{2(L+1)}(m-n)}-\\\frac{\sin
\frac{\pi}{2}(m+n)}{\sin \frac{\pi}{2(L+1)}(m+n)}\Big]. 
\label{c2_obc}
\end{multline}
For free-fermionic models the reduced density matrix for $A$ can be written as 
\begin{equation}
\rho_A=\frac{1}{Z}\exp\left(-{\mathcal H}_E\right)
\label{rho_A}
\end{equation}
where $Z$ ensures the normalization $\textrm{Tr}\rho_A=1$, and ${\mathcal H}_E$ is 
the so-called entanglement Hamiltonian. The spectrum of the reduced density matrix $\rho_A$ 
in Eq.~\eqref{rho_A}, which is expressed in the free-fermionic variables $c_i$, coincides 
with that of the reduced density matrix of the same block $A$ expressed in the original spin 
variables $\sigma^{x,y,z}_i$. Since the Jordan-Wigner transformation Eq.~\eqref{j-wigner} 
is non-local, this is a non-trivial fact, and it rests on subsystem $A$ being a 
single interval. In fact,  it does not remain true for two (or many) disjoint 
intervals~\cite{fagotti-2010,alba-2010,alba-2011,igloi-2010,fagotti-2011,fagotti-2012}.

The spectrum of ${\mathcal H}_E$ (single-particle entanglement spectrum) 
is of the free-type, reflecting the original Hamiltonian Eq.~\eqref{xx_fer} 
being quadratic, and its single-particle levels $\epsilon_l$ ($l$ 
being an arbitrary label) are obtained from the eigenvalues $\zeta_l$ 
of ${\mathbb G}_{m,n}$ (cf. Eq.~\eqref{c2}) as 
\begin{equation}
\epsilon_l=\log\left[\frac{1-\zeta_l}{\zeta_l}\right]. 
\end{equation}
Finally, the von Neumann entropy $S_A$ of $A$ is obtained as 
\begin{equation}
S_A=\sum\limits_{l=1}^{L_A}\left[\log(1+e^{-\epsilon_l})+
\frac{\epsilon_l}{1+e^{\epsilon_l}}\right]. 
\end{equation}
Equivalently, in terms of $\zeta_k$ one can write 
\begin{equation}
S_A=\sum\limits_{l=1}^{L_A}\left[\zeta_l\log\zeta_l-(1-\zeta_l)
\log(1-\zeta_l)\right].
\label{vn-en}
\end{equation}
It is noteworthy that the term in the sum in Eq.~\eqref{vn-en} has a 
maximum at $\zeta_l=1/2$, whereas it is vanishing for $\zeta_l=0,1$.

\begin{figure}
\includegraphics[width=.9\linewidth]{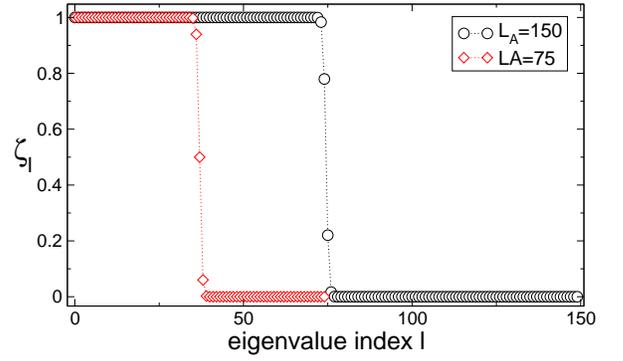}
\caption{ Ground-state single-particle entanglement spectrum (ES) 
 levels for the $XX$ chain. Symbols are exact results 
 for a chain of length $L=300$. The ES levels are for a  subsystem 
 with $L_A=75$ (rhombi) and $L_A=150$ (circles). Only few ES levels 
 (with $\zeta_l\ne 0,1$) contribute to the entanglement entropy.
}
\label{fig12:gs_sES}
\end{figure}

An example of a single-particle entanglement spectrum is shown in 
Fig.~\ref{fig12:gs_sES} plotting the eigenvalues $\zeta_l$ for a periodic 
$XX$ chain with $L=300$ and block $A$ with $L_A=75$ and $L_A=150$ (rhombi 
and circles in the figure, respectively). Clearly, a large fraction of the 
spectrum (levels with $\zeta_l=0,1$) does not contribute to the entanglement 
entropy (cf. Eq.~\eqref{vn-en}).

\section{Entanglement entropies after a geometric quench}
\label{ent_free_geo}

In this section we illustrate the calculation of the entanglement entropy 
at {\it any} time after a generic (i.e., with arbitrary aspect ratio 
$\omega$, see Fig.~\ref{fig00:quench_prot}) geometric quench in the $XX$ 
chain. Similar results can be obtained for the $1D$ $XY$ model (see 
Refs.~\onlinecite{barouch-1970,barouch-1971,barouch-1971a,mccoy-1971})  
or the transverse-field Ising chain (see Refs~\onlinecite{perk-1977,perk-2009} 
and references therein). For simplicity, here  we consider the 
situation in which both the initial chain $A$ (see Fig.~\ref{fig00:quench_prot}) 
and the final one have periodic boundary conditions. Notice that this 
implies that the quench protocol (see Fig.~\ref{fig00:quench_prot}) involves 
a ``cut and glue'' step.  

At $t=0$ the initial state of the $XX$ chain is obtained by gluing together the 
zero-magnetization ground state of a chain of length $L_A$ with a 
fully-polarized state $|F \rangle\equiv |\uparrow\uparrow\uparrow\cdots
\uparrow\rangle$ of length $L_B\equiv L-L_A$. The two-point correlation 
matrix ${\mathbb G}_{m,n}(t)$ at $t=0$ is given as 
\begin{equation}
{\mathbb G}_{m,n}(0)=\left\{\begin{array}{ccc}
{\mathbb G}^{(\textrm{\textrm{init}})}_{m,n} & \textrm{if} & (m,n)\in A\\
\delta_{m,n} & \textrm{otherwise} &
\end{array}\right.
\end{equation}
with ${\mathbb G}^{(\textrm{init})}_{m,n}$ the $t=0$ correlation function 
in part $A$ of the chain, and $\delta_{m,n}$ the Kronecker delta. 
Here we choose ${\mathbb G}^{(\textrm{init})}_{m,n}=
{\mathbb G}_{m,n}^{(\textrm{pbc})}$ (Eq.~\eqref{c2_pbc_sum} after replacing $L\to L_A$). 
At $t>0$ after the quench ${\mathbb G}_{m,n}(t)$ is obtained as follows. One first defines 
$U_{kj}\equiv\sum_m R_{km}e^{iE_m t}(R^\dagger)_{mj}$, where $R_{kj}$ 
is constructed as $R_{kj}\equiv\langle 0|c_j|v_{k}\rangle$ and $E_{m}$ 
is given by Eq.~\eqref{disp_per}. One then has 
\begin{equation}
{\mathbb G}(t)=U^\dagger {\mathbb G}(0)U\,.
\label{xx_evol}
\end{equation}
The explicit expression after performing the matrix multiplications 
in Eq.~\eqref{xx_evol} reads  
\begin{multline}
{\mathbb G}_{m,n}(t)=\sum\limits_{k,k'}e^{-2\pi i
\frac{k}{L}m+2\pi i\frac{k'}{L}n+i(E_{k'}-E_k)t}
\Big[\frac{1}{L}\delta_{kk'}-\\\frac{1}{L^2}
\frac{1}{L_A}\sum\limits_{r=-k_F}^{k_F}\frac{1-
e^{2\pi i\omega k}}{1-e^{i\frac{2\pi}{L_A}
(r+\omega k)}}\times\frac{1-e^{-2\pi 
i\omega k'}}{1-e^{-i\frac{2\pi}{L_A}
(r+\omega k')}}\Big].
\label{corr}
\end{multline}
It is convenient to define the matrix ${\mathbb F}_{r,m}(t)$ as 
\begin{equation}
{\mathbb F}_{r,m}(t)\equiv\frac{1}{L}\sum\limits_{k}e^{-2\pi i
\frac{k}{L}m-iE_kt}\frac{1-e^{2\pi i\omega k}}{1-
e^{i\frac{2\pi}{L_A}(r+\omega k)}}.
\end{equation}
Thus one can rewrite Eq.~\eqref{corr} in the form  
\begin{equation}
\label{corr_t}
{\mathbb G}_{m,n}(t)=\delta_{m,n}-\frac{1}{L_A}\sum\limits_{r=-k_F}^{k_F}
{\mathbb F}_{m,r}(t){\mathbb F}^*_{n,r}(t).
\end{equation}
Finally, the entanglement entropy for part $A$ of the chain after 
the geometric quench is obtained from the eigenvalues of Eq.~\eqref{corr_t} 
restricted to $A$, using Eq.~\eqref{vn-en}.

\section{Dynamics after quenching to the infinite chain}
\label{inf-quench}

In this section we focus on the entanglement dynamics after a geometric 
quench in the limit $\omega\to 0$ (quench to the infinite chain, cf. 
Fig.~\ref{fig00:quench_prot}). Notice that the limit $\omega\to 0$ is 
taken at fixed finite $L_A$.  

The time evolved correlation function ${\mathbb G}_{m,n}(t)$ has the 
same form as in Eq.~\eqref{corr_t} after redefining ${\mathbb F}_{m,r}(t)$ 
as  
\begin{equation}
{\mathbb F}_{m,r}(t)=\frac{1}{2\pi}\int_0^{2\pi}dk 
\frac{1-e^{iL_Ak}}{1-e^{i\frac{2\pi r}{L_A}+ik }
}e^{-imk+it\cos k}\,.
\label{inf_F}
\end{equation}
One should observe that in the limit $L_A\gg 1$, the second term in 
the numerator in Eq.~\eqref{inf_F} is highly oscillating and one can write 
\begin{multline}
{\mathbb F}_{m,k_s}(t)\xrightarrow{L_A\gg 1} \frac{1}{2\pi}{\mathcal P}
\int_0^{2\pi}dk\frac{e^{-imk+it\cos k}}{1-e^{i(k+k_s)}}
+\\\frac{1}{2}e^{i mk_+it\cos k_s}
\label{inf_F_delta}
\end{multline}
where ${\mathcal P}$ denotes the Cauchy principal value of the 
integrand and we introduced $k_s\equiv 2\pi r/L_A$.  
The approximation Eq.~\eqref{inf_F_delta} holds provided that $m\ll L_A$, 
$t\ll L_A$, and $k_s\ne 0$.

A numerically more convenient expression is obtained writing  
\begin{equation}
{\mathbb F}_{m,k_s}(t)=\sum\limits_{p=0}^{L_A}e^{ipk_s+i(m-p)
\frac{\pi}{2}}J_{p-m}(2t), 
\label{bessel_corr}
\end{equation}
where $J_q(x)$ is the modified Bessel function. In the limit $L_A\to\infty$, 
using Eq.~\eqref{corr_t}, $k_s$ becomes a continuous variable and  one can 
write ${\mathbb G}_{m,n}(t)$ as   
\begin{equation}
{\mathbb G}_{m,n}(t)=\delta_{mn}-\frac{1}{2\pi}\int_{-\frac{\pi}{2}}^{
\frac{\pi}{2}}dk_s\, {\mathbb F}_{m,k_s}(t){\mathbb F}^*_{n,k_s}(t). 
\label{step}
\end{equation}
After using Eq.~\eqref{bessel_corr} and performing explicitly the 
integration in Eq.~\eqref{step}, ${\mathbb G}_{m,n}(t)$ reads 
\begin{multline}
\label{corr_bess}
{\mathbb G}_{m,n}(t)=\delta_{mn}-\\\sum\limits_{p,q=0}^{L_A-1}
\frac{\sin[\frac{1}{2}\pi(p-q)]}{\pi(p-q)}J_{p-m}(2t)J^*_{q-n}(2t)
i^{m-p-n+q}\,.
\end{multline}
Notice that we keep $L_A$ finite in Eq.~\eqref{corr_bess}, although  
formally the limit $L_A\to\infty$ is taken in Eq.~\eqref{step}. It is 
convenient to redefine $s=(p+q)/2$ and $d=(p-q)/2$ obtaining 
\begin{multline}
{\mathbb G}_{m,n}(t)=\delta_{mn}-\\\sum\limits_{d=-L_A/2}^{L_A/2}
\sum\limits_{s=|d|}^{L_A-|d|}\frac{\sin\pi d}{2\pi d}J_{s+d-m}
(2t)J^*_{s-d-n}(2t)i^{m-n-2d}\,.
\label{corr_bess1}
\end{multline}
Since $\sin(\pi d)/d$ is highly oscillating, in practice in Eq.~\eqref{corr_bess} 
it is possible to restrict the summation to the first few values of $d$. It is also 
worth reminding that $J_q(x)$ are exponentially vanishing at $|q|> x$, implying in 
Eq.~\eqref{corr_bess1} that the contribution of the sum vanishes  at 
$|s+d-m|/2>t$ and $|s-d-n|/2>t$, which is a manifestation  of the 
Lieb-Robinson bound~\cite{lieb-1972} in free models.

\section{Some remarkable scaling properties at large times}
\label{large-t}

In this section we derive the scaling form Eq.~\eqref{large-t_ent} for the  
von Neumann entropy at large times $t\to\infty$ after the quench from a finite  
to the infinite chain (cf. Appendix~\ref{inf-quench}). 

The main ingredient of the calculation is the asymptotic behavior at 
$t\to\infty$ of the matrices ${\mathbb F}_{m,r}(t)$ (cf. Eq.~\eqref{inf_F}). 
This is obtained using the stationary phase method~\cite{erdelyi-1956}. 
The large-time behavior of ${\mathbb F}_{m,r}$ depends crucially on the 
parity of $L_A$ and on the value of $r$. 

In particular, for $r=0$, irrespective of the parity of $L_A$, one 
obtains   
\begin{equation}
{\mathbb F}_{m,0}(t)\approx\frac{\sqrt{2}}{\sqrt{\pi t}}\left[
\frac{(-1)^m}{2}e^{it-i\frac{\pi}{4}}+L_Ae^{-it+i\frac{\pi}{4}}\right],
\label{odd_z}
\end{equation}
where we neglect terms ${\mathcal O}(t^{-1})$. For generic $r\ne 0$ and 
odd $L_A$ one has  
\begin{equation}
{\mathbb F}_{m,r}(t)\approx(-1)^m\frac{e^{it}}{\sqrt{4\pi t}}\Big[(1-i)-
(1+i)\frac{\sin k_s}{1+\cos k_s}\Big]
\label{odd_nz}
\end{equation}
with $k_s$ as in Eq.~\eqref{inf_F}. Finally, for $r\ne 0$ and even $L_A$ 
the result reads 
\begin{multline}
{\mathbb F}_{m,r}(t)\approx\frac{L_A(-1)^m}{\sqrt{8\pi t^3}}
\Big[i^m(2m-L_A)\cos\Big(t-m\frac{\pi}{2}+\frac{\pi}{4}\Big)\\
+\frac{1}{2}\Big(2 m\tan\frac{k_s}{2}-L_A\tan\frac{k_s}{2}+i\sec^2
\frac{k_s}{2}\Big)e^{it-i\frac{\pi}{4}}-\\(-1)^m\frac{1}{4}
\textrm{cosec}^2\frac{k_s}{2}\Big(2i+(L_A-2m)\sin k_s\Big)
e^{-it+i\frac{\pi}{4}}\Big].
\label{even_nz}
\end{multline}
Notice that Eq.~\eqref{even_nz} gives ${\mathbb F}_{m,r}(t)\sim t^{-3/2}$, which is 
subleading compared to Eqs.~\eqref{odd_z} and \eqref{odd_nz}.

The corresponding asymptotic expansion for ${\mathbb G}_{m,n}(t)$ is straightforward, 
substituting Eqs.~\eqref{odd_z}-\eqref{even_nz} into Eq.~\eqref{corr_t}. We start 
discussing the case with $L_A$ odd. The result reads   
\begin{multline}
{\mathbb G}_{m,n}(t)=\delta_{m,n}-\frac{(-1)^{m+n}}{\pi L_A t}
\sum\limits_{r=-k_F}^{k_F}\frac{1}
{1+\cos\frac{2\pi r}{L_A}}\\
+i\frac{e^{-2it}[(-1)^m e^{2it}+iL_A][i(-1)^n+
e^{2it}L_A]}{2\pi L_A t}.
\label{corr_asy}
\end{multline}
It is convenient to rewrite Eq.~\eqref{corr_asy} as 
\begin{multline}
{\mathbb G}_{m,n}(t)=\delta_{m,n}+\frac{1}{t}u_mu_n [A(-1)^{m+n}+
\\B(-1)^me^{2it}+
C(-1)^ne^{-2it}+D],
\label{corr_asy1}
\end{multline}
introducing the constants $A,B,C,D$ as  
\begin{align}
& A\equiv -\frac{1}{\pi L_A }
\sum\limits_{r=-k_F}^{k_F}
\frac{1}{1+\cos\frac{2\pi r}{L_A}}-\frac{1}{2\pi L_A }\\
& B=C^*\equiv\frac{i}{2\pi }\\
& D=-\frac{L_A}{2\pi }.
\end{align}
In Eq.~\eqref{corr_asy1} we defined $u_m$ as the vector of length $L_A$ with 
unit entries, i.e., $u_m\equiv (1,1,\dots,1)$. The eigenvalues of ${\mathbb G}_{m,n}(t)$ 
can be calculated analytically. It turns out that ${\mathbb G}$ has $L_A-2$ unit 
eigenvalues. Only two eigenvalues contribute non trivially to the entanglement 
entropy, which are given as 
\begin{equation}
\label{eig_lt}
\zeta_{\pm}=1+\frac{AL_A+DL_A\pm L_A
\sqrt{A^2+4BC-2AD+D^2}}{t}\,.
\end{equation}
Using Eq.~\eqref{vn-en} and Eq.~\eqref{eig_lt}, and considering the limit $1\ll 
L_A\ll t$, one obtains that the entropy is a scaling function of $t/L_A^2$, and 
can be given as
\begin{equation}
S_A(t)\approx\frac{L_A^2}{2\pi t}\left[1-\log 
\frac{L_A^2}{2\pi t}\right]\,.
\label{odd_asy}
\end{equation}
Clearly, since $S_A(t)>0\,\forall t$, Eq.~\eqref{odd_asy} is 
valid only at $t\gg L_A^2$. 

We now turn to the case of $L_A$ even. The 
correlation matrix ${\mathbb G}_{m,n}(t)$, keeping only 
terms ${\mathcal O}(t^{-1})$ is given as 
\begin{multline}
{\mathbb G}_{m,n}(t)=\delta_{m,n}+\frac{u_mu_n}{2\pi L_A t}\times\\
[i(-1)^m -e^{-2it}L_A][i(-1)^ne^{-2it}+L_A].
\end{multline}
A similar analysis as for odd $L_A$ gives the entanglement entropy 
as
\begin{multline}
S_A(t)=-\left[1-\frac{1+L_A^2}{2\pi t}\right]\log\left[
1-\frac{1+L_A^2}{2\pi t}\right]\\-\frac{1+L_A^2}{2\pi t}\log
\frac{1+L_A^2}{2\pi t}.
\label{step1}
\end{multline}
Notice that, as expected, in the limit $1\ll L_A$ Eq.~\eqref{step1} 
reduces to Eq.~\eqref{odd_asy}. 
%


\bibliography{references}
\end{document}